%% file: Main.tex
\documentclass[11pt]{article}
\usepackage{epsfig}
\usepackage{amsmath}
\usepackage{eucal}
\usepackage[small,hang]{caption}
\usepackage{cite}
\usepackage{bm}
\newfont{\CLscriptb}{clscriptb10 scaled 1100}
\newcommand{\GaSub}{V}%
\newcommand{\CLrd}{{\rm d}}%
\newcommand{\CLrD}{{\rm D}}%
\newcommand{\CLri}{\mathrm{i}}%
\newcommand{\CLre}{\mathrm{e}}%
\newcommand{\CLrT}{\mathrm{T}}%
\newcommand{\CLrL}{\mathrm{L}}%
\DeclareMathOperator{\CLtr}{tr}%
\DeclareMathOperator{\CLdim}{dim}%
\newcommand{\alphaS}{\alpha_\mathrm{s}}%
\newcommand{\CLxB}{x_\mathrm{B}}%
\newcommand{\CLpBra}{\langle p|}%
\newcommand{\CLpKet}{|p\rangle}%
\newcommand{\CLscL}{\text{\CLscriptb L}}
\newcommand{\CLscD}{\text{\CLscriptb D}}
\newcommand{\CLnumtextS}[1]{\ifcase#1 zero\or one\or two\or three\or four\or
  five\or six\or seven\or eight\or nine\or ten\or eleven\or twelve%
  \else #1\fi}%
\numberwithin{equation}{section}%
\begin{document}
\begin{titlepage}
\hspace*{\fill}
\begin{minipage}[t]{3.5cm}
  DESY--99--118\\
  MZ--TH/99--33\\
  hep-ph/9908411
\end{minipage}
\vspace*{2.cm}
\begin{center}
  \begin{LARGE}
    \textbf{Factorization of Twist-Four Gluon Operator Contributions}\\
  \end{LARGE}
  \renewcommand{\thefootnote}{\fnsymbol{footnote}}
  \footnotetext[0]
  {Supported by the TMR Network ``QCD and Deep Structure of Elementary
    Particles''. One of us (C.B.) is supported by
    \textit{Graduiertenkolleg Theoretische Elementarteilchenphysik}.}
  \renewcommand{\thefootnote}{\arabic{footnote}}
  \setcounter{footnote}{0}
  \vspace{2.5cm}
  \begin{Large}
    {Jochen Bartels, Claas Bontus}\\
  \end{Large}
  \vspace{0.3cm}
  \textit{II.\ Institut f\"ur Theoretische Physik,
    Universit\"at Hamburg, Luruper Chaussee 149, \\
       D-22761 Hamburg\,\footnote{email:
        bartels@x4u2.desy.de $\quad$ claas.bontus@desy.de}
    }\\
  \vspace{1cm}
  \begin{Large}
    {Hubert Spiesberger}\\
  \end{Large}
  \vspace{0.3cm}
  \textit{Institut f\"ur Physik, WA ThEP,
    Johannes-Gutenberg-Universit\"at Mainz,\\
       D-55099 Mainz\,\footnote{email:
        hspiesb@thep.physik.uni-mainz.de}
    }\\
\end{center}
\vspace*{2.5cm}
\begin{quotation}
  \noindent
  We consider diagrams with up to four $t$-channel gluons in order to
  specify gluonic twist-four contributions to deep inelastic structure
  functions.  This enables us to extend the method developed by
  R.K.\,Ellis, W.\,Furmanski, and R.\,Petronzio (EFP) \cite{biblio:EFP}
  to the gluonic case. The method is based on low-order Feynman diagrams
  in combination with a dimensional analysis. It results in explicitly
  gauge invariant expressions for the factorization of twist-four
  gluon-operator matrix elements and the corresponding coefficient
  functions.
\end{quotation}
\vfill
\end{titlepage}
%
%
\input{intro.tx}
\input{foundation.tx}
\input{ward.tx}
\input{gluonops.tx}
\input{dimana.tx}
\input{facres.tx}
\input{discussion.tx}
\input{conclu.tx}
%

\input{biblio.tx}
\end{document}

%% file: intro.tx
\section{Introduction}
The currently available experimental data on deep inelastic structure functions
covers a large kinematical regime with high precision measurements. This
provides an interesting challenge for theoretical physics when it comes to
describing this data in the low-$Q^2$ domain.  At sufficiently high $Q^2$, a
perturbative approach is justified and within the \textit{classical} DGLAP
approach next-to-leading-order (NLO) fits are still the state-of-the-art.
Although some authors \cite{biblio:GRV98} have been using NLO DGLAP down to
rather low values of $Q^2$, it becomes more and more obvious
\cite{biblio:Caldwell} that such an approach is not sufficient for a precise
description of deep inelastic structure function data.  Missing are not
primarily the sub-leading terms in powers of $\alphaS$, but corrections which
are proportional to the reciprocal value of the photon virtuality $Q^2$, viz.\ 
higher-twist terms. This statement gets some support from the experimental
observation that there is a rather large number of events in which a
longitudinally polarized photon scatters off a proton diffractively and results
in a vector meson. Since this process has been shown to be calculable
perturbatively and to belong to higher-twist
\cite{biblio:Ryskin,biblio:BFGMS,biblio:CFS,biblio:MRT}, one should expect a
sizeable amount of higher-twist contributions to deep inelastic structure
functions in the same kinematical regime. However, it has been shown recently
that the situation is more complicated \cite{biblio:BarBon}: one gets
contributions with different signs which might lead to strong cancellations and
an estimate of the amount of inclusive higher-twist contributions based
exclusively on diffractive vector meson production is not sufficiently
reliable.  Therefore a systematic study of higher-twist is not only worthwhile
but necessary.  This is even more true, since at least some of the twist-four
contributions, after removing the explicit $1/Q^2$-dependence, are known to
evolve faster than the leading-twist contributions, and to increase faster with
decreasing $\CLxB$.

Most computations based on twist-four operators for the inclusive process have
been published in the early 1980s. S.\,Gottlieb \cite{biblio:Gottlieb}
classified all twist-four operators and computed some of their anomalous
dimensions.  The anomalous dimensions of four-quark operators and of flavor
octet, three-body operators were obtained by M.\,Okawa \cite{biblio:Okawa}, and
the Wilson coefficient functions of four-quark operators by S.P.\,Luttrell
et.al.\ \cite{biblio:Luttrell}. Calculations based on the complete set of
twist-four operators introduced in \cite{biblio:Gottlieb} are difficult since
this set is over-determined: different operators are related by the equations
of motion. This deficiency was remedied by R.L.\,Jaffe and M.\,Soldate
\cite{biblio:JaffeSol,biblio:Jaffe} who introduced a special basis, which they
called \textit{canonical}. They showed that this basis is adequate for the
necessary twist-four perturbative calculations and calculated the operators'
tree-level coefficient functions.

These approaches are all based on the operator product expansion (OPE) and make
use of the fact that the hadronic tensor $W^{\alpha\beta}$ factorizes into two
parts, viz.\ coefficient functions $C_i$ and parton correlation functions $f_i$
which depend on a set of light-cone momenta $\{x\}$:
\begin{equation}
  W^{\alpha\beta}=\sum_{\substack{\tau=2 \\ \tau\text{ even}}}
  \left(\frac{\Lambda^2}{Q^2}\right)^{\frac{\tau-2}2}
  \sum_i \int\!\!\CLrd\{x\}\,
  C_{i,\tau}^{\alpha\beta}(\{x\})f_{i,\tau}(\{x\}) \,,
\end{equation}
where $\tau$ corresponds to the twist of the quantities and we have dropped
details like the dependence on logarithms of $Q^2$. On the other hand,
R.K.\,Ellis, W.\,Furmanski and R.\,Petronzio (EFP) \cite{biblio:EFP} followed a
more intuitive approach and computed $1/Q^2$-corrections to the conventional
leading-twist Born diagram starting from leading-order Feynman diagrams.  Using
special techniques to classify the spinor, color and momentum structure, the
authors could perform the factorization of quark-coefficient functions and
operators and reduced the number of contributing terms with the help of
equations of motion.

The computations in \cite{biblio:EFP} can coarsely be divided into two steps.
In the first step the authors considered the relevant diagrams with four
$t$-channel partons at the Born level and performed the factorization of the
coefficient functions and matrix elements of the relevant operators. In detail,
these operators are (we are ignoring terms which involve a matrix
$\gamma_5$)\footnote{Cf.\ Eqs.\,(2.45)-(2.50) and (5.23)-(5.24) in
  \cite{biblio:EFP}. The calculations are performed in the axial gauge with the
  gauge vector $n_{\mu}$. Since we are interested in the general structure of
  the results, we postpone the definition of various quantities until later
  sections.}:
\begin{align}
  \label{intro:EFPTwoQuarkOp1}
  A^\rho(\lambda)&=\frac14 \CLpBra\CLrT
    \left\{\bar\psi(0)\gamma^\rho\psi(\lambda n)\right\} \CLpKet \,,\\
  \label{intro:EFPTwoQuarkOp2}
  B^{\rho\mu}(\eta,\lambda)&=\frac14 \CLpBra\CLrT
    \left\{\bar\psi(0)\gamma^\rho\CLrD^\mu(\eta n)\psi(\lambda n)\right\}
    \CLpKet \,,\\
  \label{intro:EFPTwoQuarkOp3}
  C^{\rho\mu\nu}(\omega,\eta,\lambda)&=\frac14 \CLpBra\CLrT
    \left\{\bar\psi(0)\gamma^\rho\CLrD^\mu(\omega n)\CLrD^\nu(\eta n)
      \psi(\lambda n)\right\} \CLpKet \\
\intertext{and}
  \label{intro:EFPFourQuarkOp1}
  Q(\omega,\eta,\lambda)&=\frac1{16}g^2 \CLpBra\CLrT
    \left\{\left[\bar\psi(0)\!{\not\!n}
        t^a\psi(\omega n)\right]
      \left[\bar\psi(\eta n)\!{\not\!n}t^a\psi(\lambda n)\right]
      \right\} \CLpKet \,.
\end{align}

In the second step, EFP used equations of motion in order to express the
contributions of the two-quark operators
\eqref{intro:EFPTwoQuarkOp1}-\eqref{intro:EFPTwoQuarkOp3} in terms of linearly
independent operators. For this they studied both the longitudinal basis (which
corresponds to the canonical basis in \cite{biblio:JaffeSol}) and introduced a
new one which they called the transverse basis. While within the longitudinal
basis the expressions turned out to be rather complicated, in the transverse
basis the twist-four contributions of the two-quark operator to the transverse
and longitudinal structure functions $F_\CLrT$ and $F_\CLrL$ can be expressed
in terms of
\begin{align}
  \label{intro:EFPcontribs1}
  \Lambda^2\bm T_1(x) &=\frac14 \int\!\frac{\CLrd\lambda}{2\pi}
    \CLre^{\CLri\lambda x}\CLpBra\CLrT
    \left\{\bar\psi(0)\gamma_\mu\!{\not\!n}\gamma_\nu
      d^{\mu\mu'} \CLrD_{\mu'}(0) d^{\nu\nu'}
      \CLrD_{\nu'}(\lambda n)\psi(\lambda n)\right\}\CLpKet \\
\intertext{and}
  \label{intro:EFPcontribs2}
  \Lambda^2\bm T_2(x_2,x_1) &=\frac14 \int\!\frac{\CLrd\lambda}{2\pi}
    \frac{\CLrd\eta}{2\pi}\CLre^{\CLri\eta(x_2-x_1)}
    \CLre^{\CLri\lambda x_1}\CLpBra\CLrT
    \left\{\bar\psi(0)\gamma_\mu\!{\not\!n}\gamma_\nu
      d^{\nu\nu'} \CLrD_{\nu'}(\eta n) d^{\mu\mu'}
      \CLrD_{\mu'}(\eta n)\psi(\lambda n)\right\}\CLpKet \,,
\end{align}
such that the twist-four contributions to $F_\CLrT$ and $F_\CLrL$ are
(we are neglecting the proton mass)
\begin{align}
  \label{intro:EFPSolutionFT}
  F_\CLrT^{\tau=4} &=\frac{\Lambda^2}{Q^2}
    \left[4\bm T_1(\CLxB) - \CLxB\int\!\!\CLrd x_1\CLrd x_2
      \frac{\delta(\CLxB-x_2)-\delta(\CLxB-x_1)}{x_2-x_1}\bm T_2(x_2,x_1)
      \right] \,,\\
  \label{intro:EFPSolutionFL}
  F_\CLrL^{\tau=4} &=4\frac{\Lambda^2}{Q^2} \bm T_1(\CLxB)
\end{align}
plus terms involving the four-quark operator \eqref{intro:EFPFourQuarkOp1}. 

EFP's results have been used as a starting point for many other studies.  For
example, introducing a quantity named \textit{special propagator}, J.\,Qiu
\cite{biblio:Qiu} could reproduce the results obtained in \cite{biblio:EFP} in
a very elegant way.  A.P.\,Bukhvostov et.al.\ \cite{biblio:Bukhvostov} define a
special class of operators which they call \textit{quasi-partonic}, formulate
the evolution equations for these operators and compute the integral kernels
for all possible $2\to2$ transitions in the one-loop approximation.

All these computations of twist-four quantities are based on quark-operators.
In the small-$\CLxB$ domain, on the other hand, we know that in the
leading-twist case the gluon-operator is dominating since it leads to the
maximum power in $1/\CLxB$.  For higher-twist, one expects that in the
small-$\CLxB$ domain it is, again, the gluon-operators which lead to the main
contributions.  A systematic analysis of twist-four gluon-operators is missing.
Such an analysis could, again, be based on the OPE\@.  For twist-four one has
to take into account contributions of two-gluon operators,
\begin{equation} \label{intro:Two_G_Op1}
  \CLtr F^{\mu_1 \alpha} \CLrD^{\mu_2}\dots \CLrD^{\mu_{n-1}}
    F^{\mu_n}_{\phantom{\mu_n}\alpha}\,
    g_{\mu_i\, \mu_j}\;+\; {\rm perm.,}
\end{equation}
as well as of four-gluon operators,
\begin{align} \label{intro:Four_G_Op1}
  &\CLtr F^{\mu_1 \alpha}\dots F^{\mu_i}_{\phantom{\mu_i}\alpha}
                  \dots F^{\mu_j \beta}\dots
                   F^{\mu_n}_{\phantom{\mu_n}\beta} \,,   \nonumber \\
  &\CLtr F^{\mu_1 \alpha} \dots F^{\mu_i \beta}
                  \dots F^{\mu_j}_{\phantom{\mu_j}\alpha}\dots
                   F^{\mu_n}_{\phantom{\mu_n}\beta} \,,   \\ 
  &\CLtr F^{\mu_1 \alpha} \dots F^{\mu_i \beta}
                  \dots F^{\mu_j}_{\phantom{\mu_j}\beta}\dots
                  F^{\mu_n}_{\phantom{\mu_n}\alpha}  \nonumber \,.
\end{align} 
Here the dots denote products of covariant derivatives. In a future analysis
one will have to compute the corresponding coefficient functions and the
anomalous dimensions. This is a demanding task since all these twist-four
gluon-operators will mix under renormalization.

Alternatively, one can make use of the more intuitive approach of EFP in order
to find expressions for the contributions of twist-four gluon-operators to
inclusive DIS structure functions.  This is what we want to discuss in this
article.  We base our computations on the diagrams in Fig.\,\ref{FactDiags1}
and consider amplitudes with two, three and four $t$-channel gluons.  The
method we are going to describe will lead to the factorization of gluonic
coefficient functions and matrix elements of gluon-operators between proton
states.  For our analysis we are neglecting the target-mass corrections and we
do not start from the classical way of defining twist as the difference of
canonical dimension and spin; instead we will search for $1/Q^2$ corrections to
the familiar leading-twist result. We will use a dimensional analysis to show
that diagrams with more than four $t$-channel gluons lead to contributions of,
at least, twist-six.  We are working in the axial gauge, and we can show that,
within this gauge, it is sufficient to restrict oneself to the diagrams in
Fig.\,\ref{FactDiags1}.  In this paper we do not compute explicit expressions
for the upper quark-loop parts in the diagrams of Fig.\,\ref{FactDiags1}.
Instead we derive some general expressions for the corresponding amplitudes and
make use only of the color-, Lorentz- and momentum structure of these
quark-loops.

In comparison to quarkonic operators considered by EFP, the case of gluonic
operators is slightly more complicated, and we will carry out only the very
first steps of the corresponding analysis. Since the photon couples to the
four-gluon state only through the quark loop, already the ``Born''
approximation will contain contributions proportional to $\log Q^2$ (i.e.,
UV-divergences). In this first part of the analysis, we will ignore logarithms
and we will not address the computation of the quark loops. The emphasis of the
present paper lies on the analysis of the color and Lorentz structure and on
the factorization of the quark loop and the proton matrix elements. We will not
yet consider the moments of the structure functions and also the definition of
an operator basis will be postponed to a future paper.

For our discussion it is useful to clarify the notion of leading-order (LO) and
next-to-leading order. The diagrams in Figs.\,\ref{FactDiags1}b and c are
suppressed by factors $g$ and $g^2$, resp., compared to the diagram in
Fig.\,\ref{FactDiags1}a. Within our derivations we will factor these constants
out and absorb them within the expressions for the lower blobs.  This way we
can treat all three quark-loop expressions as if they were of the same order in
$\alphaS$.  On the other hand, computing these quark-loop expressions one will
get terms proportional to logarithms in $Q^2$ and terms which are just
constants.  Terms which are proportional to logarithms correspond to LO
quark-coefficient functions convoluted by Altarelli-Parisi splitting functions,
while we denote the constants as NLO gluon-coefficient functions.  For this
first analysis we restrict ourselves to quark-loop diagrams which are
one-particle irreducible (1PI). The three- and four-gluon vertices in
Fig.\,\ref{ThreeFourGVert1} have, therefore, to be considered as if they were
absorbed by the lower blobs and are part of the evolution.

For our computations it is convenient to use the axial gauge, i.e.\ we are
introducing a light-like vector $n^\mu$ such that
\begin{equation}
  n_\mu A^\mu_a = 0 \,.
\end{equation}
We are using the same vector $n$ within the decomposition of momentum vectors
into Sudakov variables:
\begin{equation} \label{intro:Sudakov1}
  k_i^\mu = x_i p^\mu + \alpha_i n^\mu + k_{i,\mathrm T}^\mu \,,
  \qquad p^2=n^2= p\cdot k_{i,\mathrm T} = n\cdot k_{i,\mathrm T} =0 \,,
  \qquad p\cdot n=1 \,.
\end{equation}
Here $p$ represents the momentum of the target hadron, which we assume to be
massless throughout this paper. For the color factors we have chosen the
following normalization of the Gell-Mann matrices $t^a$:
\begin{equation}
      \CLtr t^a t^b = \frac12 \delta^{ab}\, , 
\end{equation}
and we set the number of colors $N_\mathrm c=3$.

The outline of this paper is as follows. In section~\ref{Sec:fou} we define our
basic quantities.  In particular we distinguish between the upper quark-loop
parts in Fig.\,\ref{FactDiags1}, $R$, and the lower blobs, $\Gamma$.  The
expressions for the quark-loops will then be expanded about the longitudinal
components of the gluon-momenta they depend on. For the derivatives which occur
in these expansions, we compute some Ward identities in section~\ref{Sec:Ward}.
In section~\ref{Sec:Ops} we show how, in the axial gauge, the
$A_a^\mu$-operators which occur in the matrix elements for the lower blobs in
Fig.\,\ref{FactDiags1} can be expressed in terms of $F_a^{\mu\nu}$-operators.
Section~\ref{Sec:DimAnalysis} is devoted to the dimensional analysis of the
quantities under consideration. This analysis leads to the justification that
we can restrict ourselves to the diagrams in Fig.\,\ref{FactDiags1} for a
consistent analysis of gluonic twist-four contributions. Since the diagrams in
Fig.\,\ref{FactDiags1} contain also leading-twist contributions and
contributions of twist larger than four, we will provide, also in
section~\ref{Sec:DimAnalysis}, a method for projecting out the twist-four
parts. Using the results of sections~\ref{Sec:fou}
through~\ref{Sec:DimAnalysis}, the twist-four contributions we are interested
in can be written in a completely gauge invariant form, which is factorized
into coefficient functions and matrix elements between proton states.  We give
these results in section~\ref{Sec:facres}, viz.\ 
Eqs.\,\eqref{facres:H1gaugeInv1}, \eqref{facres:H2gaugeInv1} and
\eqref{facres:H3gaugeInv1}.  Subsection~\ref{Sec:facres:FGDiscussion} contains
a comparison between our results and those obtained by EFP\@.  Finally, in the
conclusions, we discuss the steps which have to be completed before our results
can be used in a phenomenological analysis.


%% file: foundation.tx
\begin{figure}
  \begin{center}
    \input{ampli1.pstex_t}
    \caption{For a systematic analysis of the twist-four contributions of
      gluonic operators it is sufficient to consider the contributions of
      the three diagrams shown here. The quark-loops symbolize the
      coupling of the gluons in all possible ways and the propagators of
      the gluon-lines belong to the lower blobs. We take all gluon momenta
      flowing upwards, i.e.\ into the quark-loops.}
    \label{FactDiags1}
  \end{center}
\end{figure}
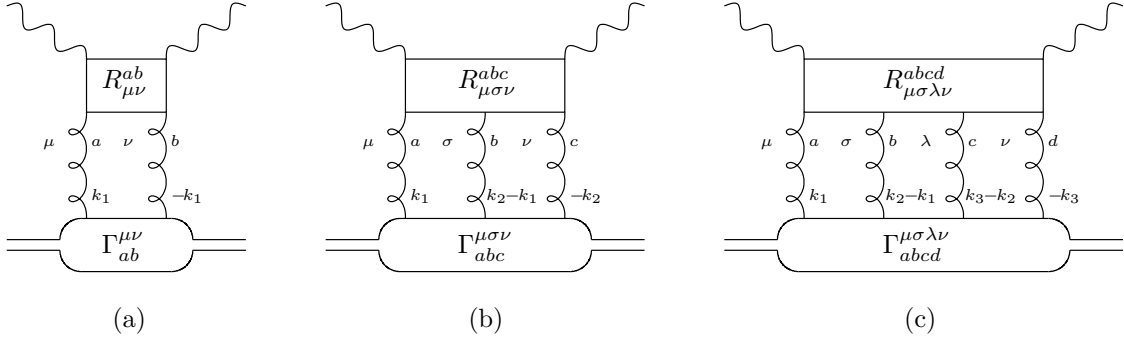
%
%
\section{Factorization} \label{Sec:fou}
As outlined in the introduction it is our aim to specify twist-four gluon
operators, write them in covariant form and perform the factorization of
coefficient functions and matrix elements of these operators between proton
states. We will argue that it is sufficient to consider the diagrams in
Fig.\,\ref{FactDiags1}. Here the upper parts represent the quark-loops with
two, three and four $t$-channel gluons. They are amputated in the gluon lines,
and we denote them by $R_{\mu\nu}^{ab}(k_1)$, $R_{\mu\sigma\nu}^{abc}(k_1,k_2)$
and $R_{\mu\sigma\lambda\nu}^{abcd}(k_1,k_2,k_3)$, resp. These quark-loops will
contribute to both NLO twist-four gluon-coefficient functions and terms which
are convolutions of LO quark-coefficient functions with Altarelli-Parisi
splitting functions. It is our aim to derive general results for the twist-four
contributions of the diagrams in Fig.\,\ref{FactDiags1} without using explicit
expressions for these quark-loops. All we need are some general properties of
the functions $R$ concerning the Lorentz-, color- and momentum-structure.

The lower parts in Fig.\,\ref{FactDiags1} are denoted by
$\Gamma^{\mu\nu}_{ab}(k_1)$, $\Gamma^{\mu\sigma\nu}_{abc}(k_1,k_2)$ and
$\Gamma^{\mu\sigma\lambda\nu}_{abcd}(k_1,k_2,k_3)$, resp., and with these
conventions the contributions of the diagrams in Fig.\,\ref{FactDiags1} are
given by
\begin{eqnarray}
  \label{fou:tildeH1}
  \tilde H_1(\CLxB,Q^2)&=&\int\!\!\frac{\CLrd^4k_1}{(2\pi)^4}
    R_{\mu\nu}^{ab}(k_1)\Gamma^{\mu\nu}_{ab}(k_1) \, , \\
  \label{fou:tildeH2}
  \tilde H_2(\CLxB,Q^2)&=&\int\!\!\frac{\CLrd^4k_1}{(2\pi)^4}
    \frac{\CLrd^4k_2}{(2\pi)^4}
    R_{\mu\sigma\nu}^{abc}(k_1,k_2)\Gamma^{\mu\sigma\nu}_{abc}(k_1,k_2) \, , \\
  \label{fou:tildeH3}
  \tilde H_3(\CLxB,Q^2)&=&\int\!\!\frac{\CLrd^4k_1}{(2\pi)^4}
    \frac{\CLrd^4k_2}{(2\pi)^4} \frac{\CLrd^4k_3}{(2\pi)^4}
    R_{\mu\sigma\lambda\nu}^{abcd}(k_1,k_2,k_3)
    \Gamma^{\mu\sigma\lambda\nu}_{abcd}(k_1,k_2,k_3) \,,
\end{eqnarray}
where we have dropped the photon indices.  As we will show in
section~\ref{Sec:DimAnalysis} expressions
\eqref{fou:tildeH1}-\eqref{fou:tildeH3} contain leading-twist and twist-four
contributions of gluonic operators as well as some parts with twist larger than
four. The twist-four parts can be projected out if one performs a dimensional
analysis. Here, the convenience of the axial gauge becomes most explicit.

The lower blobs in Fig.\,\ref{FactDiags1} can be written as matrix elements of
gluon-operators between proton states:
\begin{align}
  \label{fou:GammaDef1a}
  \Gamma^{\mu\nu}_{ab}(k_1) &= \int\!\!\CLrd^4z_1 \, \CLre^{\CLri k_1\cdot z_1}
    \CLpBra\CLrT\{A_b^\nu(0)A_a^\mu(z_1)\}\CLpKet \, ,\\
  \label{fou:GammaDef1b}
  \Gamma^{\mu\sigma\nu}_{abc}(k_1,k_2) &= \int\!\!\CLrd^4z_1\,\CLrd^4z_2 \,
    \CLre^{\CLri k_1\cdot z_1}\CLre^{\CLri(k_2-k_1)\cdot z_2}
    \CLpBra\CLrT\{A_c^\nu(0)A_b^\sigma(z_2)A_a^\mu(z_1)\}\CLpKet \, ,\\
  \label{fou:GammaDef1c}
  \Gamma^{\mu\sigma\lambda\nu}_{abcd}(k_1,k_2,k_3) &= \int\!\!\CLrd^4z_1\,
    \CLrd^4z_2\,\CLrd^4z_3\, 
    \CLre^{\CLri k_1\cdot z_1}\CLre^{\CLri(k_2-k_1)\cdot z_2}
    \CLre^{\CLri(k_3-k_2)\cdot z_3}
    \CLpBra\CLrT\{A_d^\nu(0)A_c^\lambda(z_3)A_b^\sigma(z_2)
    A_a^\mu(z_1)\}\CLpKet \,.
\end{align}
Here $\CLrT$ denotes the time ordered product, and expressions
\eqref{fou:GammaDef1a}-\eqref{fou:GammaDef1c} contain the propagators of the
gluon lines in Fig.\,\ref{FactDiags1}.

In order to perform the factorization procedure with expressions
\eqref{fou:tildeH1}-\eqref{fou:tildeH3} it is convenient to expand the
expressions for the quark-loops about the longitudinal components of the
gluon-momenta (cf.\,\eqref{intro:Sudakov1}):
\begin{align}
  \label{fou:Rexp1}
  &R_{\mu\nu}^{ab}(k_1) = R_{\mu\nu}^{ab}(x_1p) +
    \frac{\partial R_{\mu\nu}^{ab}(x_1p)}{\partial k_1^\sigma}(k_1-x_1p)^\sigma
    + \frac12\frac{\partial^2R_{\mu\nu}^{ab}(x_1p)}{\partial k_1^\sigma
    \partial k_1^\lambda}(k_1-x_1p)^\sigma(k_1-x_1p)^\lambda + \cdots \,,\\
  \label{fou:Rexp2}
  &R_{\mu\sigma\nu}^{abc}(k_1,k_2) = R_{\mu\sigma\nu}^{abc}(x_1p,x_2p) +
    \sum_{i=1}^2 \frac{\partial R_{\mu\sigma\nu}^{abc}(x_1p,x_2p)}
    {\partial k_i^\lambda}(k_i-x_ip)^\lambda + \cdots \, , \\
  \label{fou:Rexp3}
  &R_{\mu\sigma\lambda\nu}^{abcd}(k_1,k_2,k_3) = 
    R_{\mu\sigma\lambda\nu}^{abcd}(x_1p,x_2p,x_3p) + \cdots \, .
\end{align}
We will justify in section~\ref{Sec:DimAnalysis} that we can neglect additional
terms in these series if we are interested only in terms up to twist-four.

Before we give the results of the factorization procedure, we compute some Ward
identities for the derivatives in Eqs.\,\eqref{fou:Rexp1} and
\eqref{fou:Rexp2}, show how the expressions for the lower blobs $\Gamma$ can be
expressed with the help of gluonic $F_a^{\mu\nu}$-operators, and perform a
dimensional analysis of the expressions \eqref{fou:tildeH1}-\eqref{fou:tildeH3}
in the next three sections.
%
%
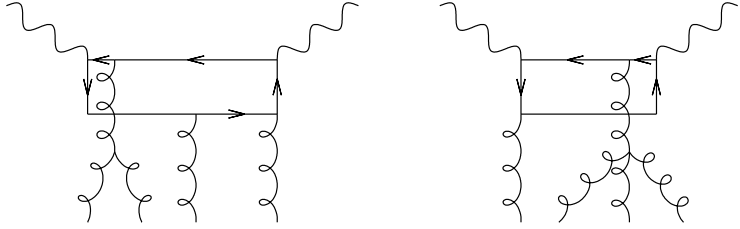
\begin{figure}
  \begin{center}
    \input{ThreeFourVert.pstex_t}
    \caption{Sample diagrams which contain three- and four-gluon vertices.}
    \label{ThreeFourGVert1}
  \end{center}
\end{figure}


%% file: ampli1.pstex_t
\begin{picture}(0,0)%
\epsfig{file=ampli1.pstex}%
\end{picture}%
\setlength{\unitlength}{3947sp}%
\begingroup\makeatletter\ifx\SetFigFont\undefined%
\gdef\SetFigFont#1#2#3#4#5{%
  \reset@font\fontsize{#1}{#2pt}%
  \fontfamily{#3}\fontseries{#4}\fontshape{#5}%
  \selectfont}%
\fi\endgroup%
\begin{picture}(7035,2060)(439,-1629)
\put(1119,-1603){\makebox(0,0)[lb]{\smash{\SetFigFont{9}{10.8}{\rmdefault}{\mddefault}{\updefault}{\small (a)}}}}
\put(3355,-1603){\makebox(0,0)[lb]{\smash{\SetFigFont{9}{10.8}{\rmdefault}{\mddefault}{\updefault}{\small (b)}}}}
\put(6093,-1603){\makebox(0,0)[lb]{\smash{\SetFigFont{9}{10.8}{\rmdefault}{\mddefault}{\updefault}{\small (c)}}}}
\put(6994,-467){\makebox(0,0)[lb]{\smash{\SetFigFont{9}{10.8}{\rmdefault}{\mddefault}{\updefault}{\tiny $d$}}}}
\put(5993,-467){\makebox(0,0)[lb]{\smash{\SetFigFont{9}{10.8}{\rmdefault}{\mddefault}{\updefault}{\tiny $b$}}}}
\put(6493,-467){\makebox(0,0)[lb]{\smash{\SetFigFont{9}{10.8}{\rmdefault}{\mddefault}{\updefault}{\tiny $c$}}}}
\put(5492,-467){\makebox(0,0)[lb]{\smash{\SetFigFont{9}{10.8}{\rmdefault}{\mddefault}{\updefault}{\tiny $a$}}}}
\put(3990,-467){\makebox(0,0)[lb]{\smash{\SetFigFont{9}{10.8}{\rmdefault}{\mddefault}{\updefault}{\tiny $c$}}}}
\put(3489,-467){\makebox(0,0)[lb]{\smash{\SetFigFont{9}{10.8}{\rmdefault}{\mddefault}{\updefault}{\tiny $b$}}}}
\put(2988,-467){\makebox(0,0)[lb]{\smash{\SetFigFont{9}{10.8}{\rmdefault}{\mddefault}{\updefault}{\tiny $a$}}}}
\put(1486,-467){\makebox(0,0)[lb]{\smash{\SetFigFont{9}{10.8}{\rmdefault}{\mddefault}{\updefault}{\tiny $b$}}}}
\put(985,-467){\makebox(0,0)[lb]{\smash{\SetFigFont{9}{10.8}{\rmdefault}{\mddefault}{\updefault}{\tiny $a$}}}}
\put(6694,-467){\makebox(0,0)[lb]{\smash{\SetFigFont{9}{10.8}{\rmdefault}{\mddefault}{\updefault}{\tiny $\nu$}}}}
\put(6193,-467){\makebox(0,0)[lb]{\smash{\SetFigFont{9}{10.8}{\rmdefault}{\mddefault}{\updefault}{\tiny $\lambda$}}}}
\put(5692,-467){\makebox(0,0)[lb]{\smash{\SetFigFont{9}{10.8}{\rmdefault}{\mddefault}{\updefault}{\tiny $\sigma$}}}}
\put(5191,-467){\makebox(0,0)[lb]{\smash{\SetFigFont{9}{10.8}{\rmdefault}{\mddefault}{\updefault}{\tiny $\mu$}}}}
\put(3689,-467){\makebox(0,0)[lb]{\smash{\SetFigFont{9}{10.8}{\rmdefault}{\mddefault}{\updefault}{\tiny $\nu$}}}}
\put(3188,-467){\makebox(0,0)[lb]{\smash{\SetFigFont{9}{10.8}{\rmdefault}{\mddefault}{\updefault}{\tiny $\sigma$}}}}
\put(2688,-467){\makebox(0,0)[lb]{\smash{\SetFigFont{9}{10.8}{\rmdefault}{\mddefault}{\updefault}{\tiny $\mu$}}}}
\put(1185,-467){\makebox(0,0)[lb]{\smash{\SetFigFont{9}{10.8}{\rmdefault}{\mddefault}{\updefault}{\tiny $\nu$}}}}
\put(685,-467){\makebox(0,0)[lb]{\smash{\SetFigFont{9}{10.8}{\rmdefault}{\mddefault}{\updefault}{\tiny $\mu$}}}}
\put(6994,-801){\makebox(0,0)[lb]{\smash{\SetFigFont{9}{10.8}{\rmdefault}{\mddefault}{\updefault}{\tiny $-\!k_3$}}}}
\put(6493,-801){\makebox(0,0)[lb]{\smash{\SetFigFont{9}{10.8}{\rmdefault}{\mddefault}{\updefault}{\tiny $\!k_3\!\!-\!\!k_2$}}}}
\put(5993,-801){\makebox(0,0)[lb]{\smash{\SetFigFont{9}{10.8}{\rmdefault}{\mddefault}{\updefault}{\tiny $\!k_2\!\!-\!\!k_1$}}}}
\put(5492,-801){\makebox(0,0)[lb]{\smash{\SetFigFont{9}{10.8}{\rmdefault}{\mddefault}{\updefault}{\tiny $k_1$}}}}
\put(3990,-801){\makebox(0,0)[lb]{\smash{\SetFigFont{9}{10.8}{\rmdefault}{\mddefault}{\updefault}{\tiny $-\!k_2$}}}}
\put(3489,-801){\makebox(0,0)[lb]{\smash{\SetFigFont{9}{10.8}{\rmdefault}{\mddefault}{\updefault}{\tiny $\!k_2\!\!-\!\!k_1$}}}}
\put(2988,-801){\makebox(0,0)[lb]{\smash{\SetFigFont{9}{10.8}{\rmdefault}{\mddefault}{\updefault}{\tiny $k_1$}}}}
\put(1486,-801){\makebox(0,0)[lb]{\smash{\SetFigFont{9}{10.8}{\rmdefault}{\mddefault}{\updefault}{\tiny $-\!k_1$}}}}
\put(985,-801){\makebox(0,0)[lb]{\smash{\SetFigFont{9}{10.8}{\rmdefault}{\mddefault}{\updefault}{\tiny $k_1$}}}}
\put(1052,-100){\makebox(0,0)[lb]{\smash{\SetFigFont{9}{10.8}{\rmdefault}{\mddefault}{\updefault}{\small $R_{\mu\nu}^{ab}$}}}}
\put(5959,-100){\makebox(0,0)[lb]{\smash{\SetFigFont{9}{10.8}{\rmdefault}{\mddefault}{\updefault}{\small $R_{\mu\sigma\lambda\nu}^{abcd}$}}}}
\put(3289,-100){\makebox(0,0)[lb]{\smash{\SetFigFont{9}{10.8}{\rmdefault}{\mddefault}{\updefault}{\small $R_{\mu\sigma\nu}^{abc}$}}}}
\put(1052,-1135){\makebox(0,0)[lb]{\smash{\SetFigFont{9}{10.8}{\rmdefault}{\mddefault}{\updefault}{\small $\Gamma_{ab}^{\mu\nu}$}}}}
\put(3289,-1135){\makebox(0,0)[lb]{\smash{\SetFigFont{9}{10.8}{\rmdefault}{\mddefault}{\updefault}{\small $\Gamma_{abc}^{\mu\sigma\nu}$}}}}
\put(5959,-1135){\makebox(0,0)[lb]{\smash{\SetFigFont{9}{10.8}{\rmdefault}{\mddefault}{\updefault}{\small $\Gamma_{abcd}^{\mu\sigma\lambda\nu}$}}}}
\end{picture}

%% file: ThreeFourVert.pstex_t
\begin{picture}(0,0)%
\epsfig{file=ThreeFourVert.pstex}%
\end{picture}%
\setlength{\unitlength}{3947sp}%
\begingroup\makeatletter\ifx\SetFigFont\undefined%
\gdef\SetFigFont#1#2#3#4#5{%
  \reset@font\fontsize{#1}{#2pt}%
  \fontfamily{#3}\fontseries{#4}\fontshape{#5}%
  \selectfont}%
\fi\endgroup%
\begin{picture}(4617,1404)(439,-973)
\end{picture}

%% file: ward.tx
\section{Ward Identities} \label{Sec:Ward}
In order to find relations for the derivatives of the quark-loops in
Eqs.\,\eqref{fou:Rexp1} and \eqref{fou:Rexp2} we will make use of Ward
identities. Although these identities can be derived diagrammatically at the
one-loop level, this is a rather tedious task. We, therefore, make a digression
into the path-integral formulation of QCD and derive these Ward identities in
the most general way.  For this, we first derive a relation for the generating
functional of perturbatively calculated vertex functions $\GaSub$ in the next
subsection.  This relation, viz.\ Eq.\,\eqref{ward:PertVertexWard1}, will then
be used to derive the desired equations in the following two subsections.

We should stress that the Ward identities which will be derived in this section
are valid for 1PI vertex functions. They can, therefore, not be applied to
expressions for quark-loops which involve three- and four-gluon vertices like
those in Fig.\,\ref{ThreeFourGVert1}.
%
%
\subsection{Generating Functional}
The generating functional for Green's functions in the axial gauge can be
written as \cite{biblio:Kummer,biblio:Bassetto}
\begin{equation} \label{ward:GenFuncGreen1}
  W[J_\mu^a,K^a] = \mathcal{N}\int\!\!\CLscD A_\mu^a\,\CLscD \lambda^a\,
  \exp\left\{ \CLri\int\!\!\CLrd^4 x \left[
    \CLscL_\mathrm{QCD} - \lambda^a\, n\cdot A^a + A_\mu^a J^\mu_a
    + \lambda^a K^a\right] \right\} \, .
\end{equation}
Here $\mathcal{N}$ is some normalization factor, $J_\mu^a$ and $K^a$ are
sources, the term $\lambda^a\, n\cdot A^a$ has been introduced in order to
define the axial gauge. The generating functional must be invariant under gauge
transformations, which for infinitesimal $\omega(x)$ can be written as
\begin{equation} \label{ward:GaugeTransf1}
  A_\mu^a\to A_\mu^a +\CLrD_\mu^{ab}(A) \omega^b= 
  A_\mu^a + \partial_\mu \omega^a - g f^{abc} A_\mu^c\omega^b \,.
\end{equation}
Inserting \eqref{ward:GaugeTransf1} into \eqref{ward:GenFuncGreen1},
substituting fields that occur as factors within the integral by functional
derivatives and making use of the arbitrariness of $\omega$, one gets
\begin{equation} \label{ward:GreenWard1}
  0= \CLrD_\mu^{ab}(J)\,
  \left( n^\mu \frac{\delta}{\CLri\delta K^b(x)} -
  J^\mu_b(x)\right)W[J_\mu^a,K^a]\,,
\end{equation}
where the covariant derivative is given by
\begin{equation}
  \CLrD_\mu^{ab}(J)= \delta^{ab}\partial_\mu -
  g f^{abc}\frac{\delta}{\CLri\delta J^\mu_c} \, .
\end{equation}
For our analysis we need the generating functional $\GaSub$ for vertex
functions. This can be obtained with the help of the following procedure.
First, define the functional $\tilde\GaSub$ via Legendre transformation of the
generating functional of the connected Green's functions, $Z$,
\begin{equation}
  \tilde\GaSub[A_\mu^a,\Lambda^a] = Z[J_\mu^a,K^a] -
  \int\!\!\CLrd^4 x\,\left[ A_\mu^aJ^\mu_a + K_a\Lambda^a\right]\,,
\end{equation}
where $Z$ is defined as
\begin{equation}
  Z[J_\mu^a,K^a] = -\CLri\ln W[J_\mu^a,K^a] \,.
\end{equation}
With these definitions we get from \eqref{ward:GreenWard1}
\begin{equation} \label{ward:VertexWard1}
  \CLrD_\mu^{ab}(A)\left( n^\mu\Lambda_b +
    \frac{\delta\tilde\GaSub}{\delta A_b^\mu}\right) = 0 \,.
\end{equation}
Now, $\GaSub$ can be obtained via subtraction
\cite{biblio:Kummer,biblio:Bassetto}:
\begin{equation}
  \GaSub[A_\mu^a,\Lambda^a] = \tilde\GaSub[A_\mu^a,\Lambda^a]
  + \int\!\!\CLrd^4 x\, \Lambda_a\,n\cdot A^a
\end{equation}
and the Ward identity \eqref{ward:VertexWard1} finally becomes
\begin{equation} \label{ward:PertVertexWard1}
  \CLrD_\mu^{ab}(A)\frac{\delta\GaSub}{\delta A_\mu^b} =0 \, .
\end{equation}
%
%
\subsection{Relations Between Two- and Three-Gluon Quark-Loops}
We are now deriving an equation which relates the derivative of the two-gluon
quark-loop $R_{\mu\nu}^{ab}$ with the three-gluon quark-loop
$R_{\mu\sigma\nu}^{abc}$. The relation between the perturbatively computed
quark-loops and the generating functional $\GaSub$ is as follows (see
Fig.\,\ref{FactDiags1} for our conventions on the gluon momenta: the $n$th
momentum 4-vector $k_n$, i.e.\ the $n$th argument of the $R$-functions, is the
sum of momenta of the first $n$ gluons):
\begin{align}
  \label{ward:PertQLdef1}
  \int\!\!\CLrd^4x_1\,\CLrd^4x_2\,
    \CLre^{\CLri k_1\cdot x_1}\CLre^{\CLri k\cdot x_2}
    \frac{\delta^2\GaSub[0]}{\delta A_a^\mu(x_1)\delta A_b^\nu(x_2)}&=
    (2\pi)^4\delta(k_1+k)R_{\mu\nu}^{ab}(k_1) \,,\\
  \label{ward:PertQLdef2}
  \int\!\!\CLrd^4x\,\CLrd^4x_1\,\CLrd^4x_2\,
    \CLre^{\CLri k_1\cdot x_1}\CLre^{\CLri (k_2-k_1)\cdot x}
    \CLre^{\CLri k\cdot x_2}
    \frac{\delta^3\GaSub[0]}{\delta A_a^\mu(x_1)\delta A_b^\sigma(x)
      \delta A_c^\nu(x_2)} &=
    (2\pi)^4\delta(k_2+k)R_{\mu\sigma\nu}^{abc}(k_1,k_2) \,,\\
  \label{ward:PertQLdef3}
  \int\!\!\CLrd^4x\,\CLrd^4x_1\,\CLrd^4x_2\,\CLrd^4x_3\,
    \CLre^{\CLri k_1\cdot x_1} \CLre^{\CLri (k_2-k_1)\cdot x}
    \CLre^{\CLri (k_3-k_2)\cdot x_2} \CLre^{\CLri k\cdot x_3}
    &\frac{\delta^4\GaSub[0]}{\delta A_a^\mu(x_1)\delta A_b^\sigma(x)
      \delta A_c^\lambda(x_2)\delta A_d^\nu(x_3)} = \notag\\
    &\quad=  (2\pi)^4\delta(k_3+k)R_{\mu\sigma\lambda\nu}^{abcd}
      (k_1,k_2,k_3) \,.
\end{align}
Relabelling Lorentz- and color indices in Eq.\,\eqref{ward:PertVertexWard1} and
differentiating the equation by $A_a^\mu(x_1)$ and $A_c^\nu(x_2)$ one gets
\begin{equation} \label{ward:help1}
  \partial^\sigma\frac{\delta^3\GaSub[0]}{\delta A_a^\mu(x_1)
    \delta A_b^\sigma(x)\delta A_c^\nu(x_2)} =
  g\left\{ f^{abb'}\delta(x-x_1)\frac{\delta^2\GaSub[0]}{\delta A_{b'}^\mu(x)
      \delta A_c^\nu(x_2)} -
    f^{bcb'}\delta(x-x_2)\frac{\delta^2\GaSub[0]}{\delta A_a^\mu(x_1)
      \delta A_{b'}^\nu(x)} \right\} \,.
\end{equation}
Now, multiplying Eq.\,\eqref{ward:help1} by appropriate exponentials,
integrating over $x$, $x_1$ and $x_2$ and keeping Eqs.\,\eqref{ward:PertQLdef1}
and \eqref{ward:PertQLdef2} in mind one ends up with
\begin{equation} \label{ward:help2}
  -\CLri\, k^\sigma R_{\mu\sigma\nu}^{abc}(k_1,k+k_1)=
  g\left\{ f^{abb'}R_{\mu\nu}^{b'c}(k+k_1)-
    f^{bcb'}R_{\mu\nu}^{ab'}(k_1)\right\} \, .
\end{equation}
The two gluons at the lower end of the quark loop must be in the color singlet.
We can, therefore, define
\begin{equation} \label{ward:twoGluonColStruk1}
  R_{\mu\nu}^{ab}(k) \equiv \delta^{ab}R_{\mu\nu}(k) \,.
\end{equation}
With this definition we can take the limit $k\to0$ in Eq.\,\eqref{ward:help2}
and finally get the desired result:
\begin{equation} \label{ward:TwoThreeRel1}
  \CLri\, gf^{abc}\frac{\partial R_{\mu\nu}(k_1)}{\partial k_1^\sigma} =
  R_{\mu\sigma\nu}^{abc}(k_1,k_1) \,.
\end{equation}
%
%
\subsection{Relations Involving Four-Gluon Quark-Loops}
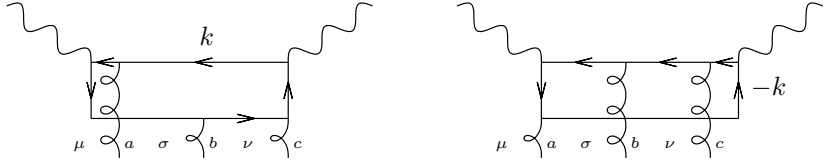
\begin{figure}[t]
  \begin{center}
    \input{ThreeQL.pstex_t}
    \caption{Two sample diagrams which enter into $R_{\mu\sigma\nu}^{abc}$ at
      leading order in $\alphaS$. With an adequate choice of the loop
      integration variable the traces over Dirac-matrices give the same (up to
      the sign) momentum dependence for both diagrams. The first diagram
      contributes with a color factor $\CLtr t^at^bt^c$ while the second one
      contributes with a color factor $\CLtr t^ct^bt^a$.} \label{fabsIllustr1}
  \end{center}
\end{figure}

Relations between derivatives of the three-gluon quark-loop
$R_{\mu\sigma\nu}^{abc}$ and the four-gluon quark-loop
$R_{\mu\sigma\lambda\nu}^{abcd}$ can be obtained in a completely analogous way.
If we differentiate \eqref{ward:PertVertexWard1} three times, multiply the
resulting equation by suitable exponentials and perform the $x_i$-integrations,
we get, similar to \eqref{ward:help2}:
\begin{align} \label{ward:help3}
  -\CLri\, k^\sigma R_{\mu\sigma\lambda\nu}^{abcd}(k_1,k_1+k,k_2+k)=
  g\Bigl\{&f^{abb'}R_{\mu\lambda\nu}^{b'cd}(k_1+k,k_2+k) \\
    &-f^{bcc'}R_{\mu\lambda\nu}^{ac'd}(k_1,k_2+k)-
    f^{bdb'}R_{\mu\lambda\nu}^{acb'}(k_1,k_2)\Bigr\} \,. \notag
\end{align}
Before we can take the limit $k\to0$ we have to explore the color structure of
$R_{\mu\sigma\nu}^{abc}$. First, consider the one-loop case.  Since
$R_{\mu\sigma\nu}^{abc}$ is the sum of all diagrams with three external gluons,
we can, in the one-loop case, always group two diagrams together. For each
diagram there is another diagram which gives the same momentum dependence (up
to the sign) but different color structure (see Fig.\,\ref{fabsIllustr1} for
illustration). Since the gluon-system must have even C-parity this color
structure is always of the form\footnote{In the case of odd C-parity the
  coupling of the gluon to quarks and anti-quarks involves a sign change. In
  this case, one therefore gets a factor
  $\CLtr(t^at^bt^c)+\CLtr(t^ct^bt^a)=\frac12 d^{abc}$.}
\begin{equation} \label{ward:ThreeGellTrace1}
  \CLtr(t^at^bt^c)-\CLtr(t^ct^bt^a)=
  \CLri\,\frac12\,f^{abc}
\end{equation}
for each pair of diagrams. We, therefore, define
\begin{equation} \label{ward:threeGluonColStruk1}
  R_{\mu\lambda\nu}^{abc}(k_1,k_2)\equiv
  gf^{abc}R_{\mu\lambda\nu}(k_1,k_2) \,,
\end{equation}
where the factor $g$ has been factored out for convenience.  We expect that
Eq.\,\eqref{ward:threeGluonColStruk1} reflects the general ansatz for the color
structure of a three-gluon system in the case of even C-parity and is valid
beyond the one-loop level. If we, now, insert \eqref{ward:threeGluonColStruk1}
into \eqref{ward:help3} and make use of the Jacobi identity,
\begin{equation} \label{ward:Jacobi1}
  f^{abc'}f^{c'cd} - f^{acc'}f^{c'bd} + f^{adc'}f^{c'bc} =0 \,,
\end{equation}
we can take the limit $k\to0$ and obtain
\begin{equation} \label{ward:ThreeFourWardId1}
  \CLri\, g^2\left\{f^{abc'}f^{c'cd}
    \frac{\partial R_{\mu\lambda\nu}(k_1,k_2)}{\partial k_1^\sigma}+
    f^{acc'}f^{c'bd}
    \frac{\partial R_{\mu\lambda\nu}(k_1,k_2)}{\partial k_2^\sigma}
  \right\} =
  R_{\mu\sigma\lambda\nu}^{abcd}(k_1,k_1,k_2) \,.
\end{equation}
%
%

For our computations we need relations for the derivatives of
$R_{\mu\lambda\nu}^{abc}$ with respect to $k_1$ or $k_2$. Since
Eq.\,\eqref{ward:ThreeFourWardId1} contains these derivatives only as a sum, we
have to project with suitable color tensors.  Now, we realize that there is
more than one color tensor which projects out the desired quantity in
Eq.\,\eqref{ward:ThreeFourWardId1}. For example both
\begin{align} \label{ward:ColProj1}
  P_1^{abcd}&\equiv \delta^{ac}\delta^{bd} \\
\intertext{and} \label{ward:ColProj2}
  P_2^{abcd}&\equiv d^{acc'}d^{c'bd}
\end{align}
project out $\partial R_{\mu\lambda\nu}/\partial k_1^\sigma$. This fact will
lead to some important restrictions for the four-gluon quark-loop
$R_{\mu\sigma\lambda\nu}^{abcd}$ (Eq.\,\eqref{ward:VanishingRelFourQL1} below).

For the remainder of this section we will restrict ourselves to the one-loop
approximation of the quark loop. In this case we can use the same arguments
that led to the definition \eqref{ward:threeGluonColStruk1}. For four gluons we
define the quantity (cf.\ \eqref{ward:ThreeGellTrace1})
\begin{align} \label{ward:dabcdDef1}
  d^{abcd}&\equiv \CLtr(t^at^bt^ct^d)+
    \CLtr(t^dt^ct^bt^a) = \notag\\
  &= -\frac14 f^{abc'}f^{c'cd} + \frac14 d^{abc'}d^{c'cd} +
    \frac16 \delta^{ab}\delta^{cd} \, .
\end{align}
We realize that there are three linearly independent tensors $d$, which can be
obtained by permutation of color indices in Eq.\,\eqref{ward:dabcdDef1} and,
therefore, write the color structure of $R_{\mu\sigma\lambda\nu}^{abcd}$ as
\begin{equation} \label{ward:FourGcolorDecomp1}
  R_{\mu\sigma\lambda\nu}^{abcd}(k_1,k_2,k_3) \equiv
  g^2\left\{ d^{abcd}R_{\mu\sigma\lambda\nu}^1(k_1,k_2,k_3) +
    d^{abdc}R_{\mu\sigma\lambda\nu}^2(k_1,k_2,k_3) +
    d^{acbd}R_{\mu\sigma\lambda\nu}^3(k_1,k_2,k_3) \right\} \,.
\end{equation}
Each diagram that contributes to $R_{\mu\sigma\lambda\nu}^{abcd}$ at the
one-loop level enters into one of the functions $R_{\mu\sigma\lambda\nu}^i$.

Inserting \eqref{ward:FourGcolorDecomp1} into \eqref{ward:ThreeFourWardId1} and
projecting with the tensors $P_1^{abcd}$ and $P_2^{abcd}$ we end up with two
equations, viz.\ 
\begin{align} \label{ward:help4}
  24\CLri\,\frac{\partial R_{\mu\lambda\nu}(k_1,k_2)}{\partial k_1^\sigma} &=
    -\frac43 R_{\mu\sigma\lambda\nu}^1(k_1,k_1,k_2) +
    \frac{32}3 R_{\mu\sigma\lambda\nu}^2(k_1,k_1,k_2) +
    \frac{32}3 R_{\mu\sigma\lambda\nu}^3(k_1,k_1,k_2) \,, \\
  20\CLri\,\frac{\partial R_{\mu\lambda\nu}(k_1,k_2)}{\partial k_1^\sigma} &=
    -\frac{40}9 R_{\mu\sigma\lambda\nu}^1(k_1,k_1,k_2) +
    \frac{50}9 R_{\mu\sigma\lambda\nu}^2(k_1,k_1,k_2) +
    \frac{50}9 R_{\mu\sigma\lambda\nu}^3(k_1,k_1,k_2) \,.
\end{align}
These two equations can simultaneously be fulfilled only if the following
equation holds\footnote{There are more tensors than those in
  Eqs.\,\eqref{ward:ColProj1} and \eqref{ward:ColProj2} which project out
  $\partial R_{\mu\lambda\nu}/\partial k_1^\sigma$ in
  \eqref{ward:ThreeFourWardId1}. Using these gives a consistent picture but no
  new results.}:
\begin{equation} \label{ward:VanishingRelFourQL1}
  R_{\mu\sigma\lambda\nu}^1(k_1,k_1,k_2) +
  R_{\mu\sigma\lambda\nu}^2(k_1,k_1,k_2) +
  R_{\mu\sigma\lambda\nu}^3(k_1,k_1,k_2) = 0 \,.
\end{equation}
Note that this equation holds only true if one of the gluon-momenta vanishes
(see Fig.\,\ref{FactDiags1} for our conventions on the gluon-momenta).
Inserting the latter equation into Eq.\,\eqref{ward:help4} we end up with the
desired results:
\begin{align}
  \label{ward:ThreeFourRel1}
  \CLri\,\frac{\partial R_{\mu\lambda\nu}^{abc}(k_1,k_2)}{\partial k_1^\sigma} 
  &=
    -\frac12 gf^{abc} R_{\mu\sigma\lambda\nu}^1(k_1,k_1,k_2) =
    -\frac12 gf^{abc} R_{\mu\lambda\sigma\nu}^3(k_1,k_2,k_2) \\
\intertext{and}
  \label{ward:ThreeFourRel2}
  \CLri\,\frac{\partial R_{\mu\sigma\nu}^{abc}(k_1,k_2)}{\partial k_2^\lambda}
  &=
    -\frac12 gf^{abc} R_{\mu\sigma\lambda\nu}^1(k_1,k_2,k_2) =
    -\frac12 gf^{abc} R_{\mu\lambda\sigma\nu}^3(k_1,k_1,k_2) \,.
\end{align}
Here we have also projected out the derivative of $R_{\mu\sigma\nu}^{abc}$ with
respect of $k_2$ in Eq.\,\eqref{ward:ThreeFourWardId1} and we have used the
fact that we have some freedom in choosing Lorentz- and color-indices and
momentum variables in defining the four-gluon quark-loop in
Eq.\,\eqref{ward:PertQLdef3}.  $R_{\mu\sigma\lambda\nu}^{abcd}$ must be
invariant under the simultaneous exchange of momentum variables, Lorentz- and
color indices. This leads to the possibility to express the derivatives of
$R_{\mu\lambda\nu}^{abc}$ in Eqs.\,\eqref{ward:ThreeFourRel1} and
\eqref{ward:ThreeFourRel2} either by $R_{\mu\sigma\lambda\nu}^1$ or by
$R_{\mu\lambda\sigma\nu}^3$. We are discussing this symmetry in more detail in
section~\ref{Sec:facres}.

Finally, differentiating Eq.\,\eqref{ward:TwoThreeRel1} by $k_1^\lambda$ and
using \eqref{ward:ThreeFourRel1} and \eqref{ward:ThreeFourRel2} we get
\begin{align} \label{ward:TwoTwiceDiff1}
  \frac{\partial R_{\mu\nu}^{ad}(k_1)}{\partial k_1^\sigma\partial k_1^\lambda}
  &= \frac12 \delta^{ad}\left\{
    R_{\mu\sigma\lambda\nu}^1(k_1,k_1,k_1)+
    R_{\mu\sigma\lambda\nu}^3(k_1,k_1,k_1) \right\}= \notag\\
  &= -\frac12 \delta^{ad} R_{\mu\sigma\lambda\nu}^2(k_1,k_1,k_1) \,,
\end{align}
where we have also made use of \eqref{ward:VanishingRelFourQL1}.


%% file: ThreeQL.pstex_t
\begin{picture}(0,0)%
\epsfig{file=ThreeQL.pstex}%
\end{picture}%
\setlength{\unitlength}{3947sp}%
\begingroup\makeatletter\ifx\SetFigFont\undefined%
\gdef\SetFigFont#1#2#3#4#5{%
  \reset@font\fontsize{#1}{#2pt}%
  \fontfamily{#3}\fontseries{#4}\fontshape{#5}%
  \selectfont}%
\fi\endgroup%
\begin{picture}(5150,998)(439,-568)
\put(875,-485){\makebox(0,0)[lb]{\smash{\SetFigFont{9}{10.8}{\rmdefault}{\mddefault}{\updefault}{\tiny $\mu$}}}}
\put(4905,-485){\makebox(0,0)[lb]{\smash{\SetFigFont{9}{10.8}{\rmdefault}{\mddefault}{\updefault}{\tiny $c$}}}}
\put(4375,-485){\makebox(0,0)[lb]{\smash{\SetFigFont{9}{10.8}{\rmdefault}{\mddefault}{\updefault}{\tiny $b$}}}}
\put(3845,-485){\makebox(0,0)[lb]{\smash{\SetFigFont{9}{10.8}{\rmdefault}{\mddefault}{\updefault}{\tiny $a$}}}}
\put(2254,-485){\makebox(0,0)[lb]{\smash{\SetFigFont{9}{10.8}{\rmdefault}{\mddefault}{\updefault}{\tiny $c$}}}}
\put(1724,-485){\makebox(0,0)[lb]{\smash{\SetFigFont{9}{10.8}{\rmdefault}{\mddefault}{\updefault}{\tiny $b$}}}}
\put(1193,-485){\makebox(0,0)[lb]{\smash{\SetFigFont{9}{10.8}{\rmdefault}{\mddefault}{\updefault}{\tiny $a$}}}}
\put(4587,-485){\makebox(0,0)[lb]{\smash{\SetFigFont{9}{10.8}{\rmdefault}{\mddefault}{\updefault}{\tiny $\nu$}}}}
\put(4057,-485){\makebox(0,0)[lb]{\smash{\SetFigFont{9}{10.8}{\rmdefault}{\mddefault}{\updefault}{\tiny $\sigma$}}}}
\put(3527,-485){\makebox(0,0)[lb]{\smash{\SetFigFont{9}{10.8}{\rmdefault}{\mddefault}{\updefault}{\tiny $\mu$}}}}
\put(1936,-485){\makebox(0,0)[lb]{\smash{\SetFigFont{9}{10.8}{\rmdefault}{\mddefault}{\updefault}{\tiny $\nu$}}}}
\put(1405,-485){\makebox(0,0)[lb]{\smash{\SetFigFont{9}{10.8}{\rmdefault}{\mddefault}{\updefault}{\tiny $\sigma$}}}}
\put(1666,164){\makebox(0,0)[lb]{\smash{\SetFigFont{12}{14.4}{\rmdefault}{\mddefault}{\updefault}{\small $k$}}}}
\put(5131,-151){\makebox(0,0)[lb]{\smash{\SetFigFont{12}{14.4}{\rmdefault}{\mddefault}{\updefault}{\small $-k$}}}}
\end{picture}

%% file: gluonops.tx
\section{Gluonic Operators} \label{Sec:Ops}
As outlined in section~\ref{Sec:fou}, the lower blobs in Fig.\,\ref{FactDiags1}
can be written as matrix-elements of gluonic operators
(Eqs.\,\eqref{fou:GammaDef1a}-\eqref{fou:GammaDef1c}). Since it is our aim to
find expressions for the gluonic twist-four contributions which are explicitly
gauge invariant, we have to find relations in which all operators consist of
gluonic operators $F^{\mu\nu}_a$ and covariant derivatives $\CLrD^\mu_{ab}$. In
this section we show how, in the axial gauge, $A^\mu_a$-operators within matrix
elements can be substituted by $F^{\mu\nu}_a$-operators. For this we make use
of the relation
\begin{equation} \label{ops:nF1}
  n_\mu F^{\mu\nu}_a = (n\cdot\partial) A^\nu_a \,.
\end{equation}

Let us, first, define\footnote{We are not writing the $p$-dependence of
  $G^{\mu\nu}_{ab}$ explicitly.}
\begin{equation} \label{ops:Def1}
  \int\!\!\CLrd^4z\,\CLrd^4z'\,
  \CLre^{\CLri k\cdot z} \CLre^{-\CLri k'\cdot z'}
  \CLpBra\CLrT\left\{ A^\nu_b(z') A^\mu_a(z) \right\} \CLpKet
  \equiv (2\pi)^4 \delta(k-k') G^{\mu\nu}_{ab}(k) \,,
\end{equation}
and examine the expression
\begin{align} \label{ops:Rel1}
  \int\!\CLrd^4z\,\CLrd^4z'\,
  &\CLre^{\CLri k\cdot z} \CLre^{-\CLri k'\cdot z'}
  \CLpBra\CLrT\left\{ (n\cdot\partial')A^\nu_b(z') 
    (n\cdot\partial)A^\mu_a(z) \right\} \CLpKet = \notag \\
  &= -\CLri\, n\cdot k\,\CLri \, n\cdot k'
    \int\!\!\CLrd^4z\,\CLrd^4z'\,
    \CLre^{\CLri k\cdot z} \CLre^{-\CLri k'\cdot z'}
    \CLpBra\CLrT\left\{ A^\nu_b(z') A^\mu_a(z) \right\} \CLpKet =
    \notag \\
  &= n\cdot k\,n\cdot k' (2\pi)^4 \delta(k-k') G^{\mu\nu}_{ab}(k) \,,
\end{align}
where we have used partial integration. If we perform the integration over $k'$
in Eq.\,\eqref{ops:Rel1}, keeping \eqref{ops:nF1} in mind and inserting
$x=n\cdot k$, we end up with
\begin{equation}
  \int\!\!\CLrd^4z\, \CLre^{\CLri k\cdot z}
  \CLpBra\CLrT\left\{ A^\nu_b(0) A^\mu_a(z) \right\} \CLpKet =
  \frac{1}{x^2} \int\!\!\CLrd^4z\, \CLre^{\CLri k\cdot z} \,
  n_\alpha n_\beta
  \CLpBra\CLrT\left\{ F^{\beta\nu}_b(0) F^{\alpha\mu}_a(z) \right\} \CLpKet \,,
\end{equation}
which is the desired result. In complete analogy, one can find relations for
matrix-elements with more than two gluon-operators.


%% file: dimana.tx
\section{Dimensional Analysis} \label{Sec:DimAnalysis}
Before we present the results of the factorization procedure we, first, perform
the dimensional analysis in this section, in order to justify that we have
restricted ourselves to diagrams with up to four $t$-channel gluons
(Fig.\,\ref{FactDiags1}) and to those terms in the expansions of the quark-loop
expressions $R$ which are given in Eqs.\,\eqref{fou:Rexp1}-\eqref{fou:Rexp3}.
By this, we also derive the projectors which are needed to obtain contributions
of definite twist from the expressions under consideration.

To understand the reasoning of the following discussion, it will be sufficient
to anticipate that after factorization the upper and lower parts of the
diagrams in Fig.\,\ref{FactDiags1} are connected by integrals over momentum
fractions $x_i$, which we can write as
\begin{align}
  \label{dimana:FacForm1a}
  H_1 &= \int\!\! \CLrd x_1\, 
    R_{\mu\nu}(x_1p)\, _2G^{\mu\nu}(x_1) \,, \\
  \label{dimana:FacForm1b}
  H_2 &= \int\!\! \CLrd x_1\CLrd x_2\, 
    R_{\mu\sigma\nu}(x_1p,x_2p)\,
    _3G^{\mu\sigma\nu}(x_1,x_2) \,, \\
  \label{dimana:FacForm1c}
  H_3 &= \int\!\! \CLrd x_1\CLrd x_2\CLrd x_3\,
    R_{\mu\sigma\lambda\nu}(x_1p,x_2p,x_3p)\,
    _4G^{\mu\sigma\lambda\nu}(x_1,x_2,x_3) \,.
\end{align}
It will also be sufficient to know that the functions $G$ are matrix elements
of gluon operators which may contain derivatives. The number of Lorentz-indices
corresponds to the number of gluon-lines plus the number of derivatives. For
convenience we are using the above relations already in this section. The
precise definition of the functions $G$ will be given below in
Eqs.\,\eqref{facres:G2matrixExp1}, \eqref{facres:G3matrixExp1} and
\eqref{facres:G41} and the functions $H_i$, which are not the same as the
functions $\tilde H_i$ in \eqref{fou:tildeH1}-\eqref{fou:tildeH3}, will be
discussed in \eqref{facres:Hidef}.

We expect an infrared cut-off to enter into the expressions for the quark-loops
only within logarithms since infrared and mass singularities are logarithmic at
most. Moreover, renormalizability guarantees that one-loop diagrams with two
external photons and any number of gluons do not contain UV divergences.
Therefore, after factorization, the only dimensional quantity the quark-loops
$R$ can depend on is $Q^2$.  On the other hand, the only dimensional quantity
the matrix-elements $G$ can depend on is $\Lambda$. If we, therefore, determine
the dimension of a particular function $R$ with Lorentz indices projected out,
we can relate this dimension to the powers of $1/Q^2$ this function must be
proportional to.

A simple analysis shows that the dimension of the quark-loops is given by the
expression
\begin{equation}
  \CLdim R = 2-B \,,
\end{equation}
where $B$ is the number of gluon-lines plus the number of derivatives:
\begin{equation}
  B = n_\mathrm{G} + n_\mathrm{D} \,.
\end{equation}
With the help of \eqref{dimana:FacForm1a}-\eqref{dimana:FacForm1c} we conclude
that
\begin{align}
  \label{dimana:DimG}
  \CLdim {_4G} = \CLdim {_3G}+1 = \CLdim {_2G} +2 \,.
\end{align}
%
%
\subsection{Two Gluons}
In order to split up $H_1$, \eqref{dimana:FacForm1a}, into leading- and
next-to-leading twist expressions, we, now, explore the tensor structure of
$_2G^{\mu\nu}$ (which is given by Eq.\,\eqref{facres:G2matrixExp1},
below)\footnote{Additionally, terms proportional to $n_\alpha
  p_\beta\varepsilon^{\mu\nu\alpha\beta}$ (in Eq.\,\eqref{dimana:Tensor2G1}) or
  $n_\alpha\varepsilon^{\mu\sigma\nu\alpha}$ (in Eq.\,\eqref{dimana:Tensor3G1})
  would be present in the case of polarized scattering or if a virtual $Z$
  boson scatters off the proton.\label{epsFootn1}}:
\begin{equation} \label{dimana:Tensor2G1}
  _2G^{\mu\nu}(x_1) = d^{\mu\nu}A_1(x_1) + n^\mu n^\nu A_2(x_1) \,.
\end{equation}
Since we are working in the axial gauge, i.e.\ $n\cdot A=0$, there can be no
terms proportional $p^\mu p^\nu$ in \eqref{dimana:Tensor2G1}
(cf.\,\eqref{fou:GammaDef1a}). The tensor
\begin{equation} \label{dimana:dmunuDef1}
  d^{\mu\nu} \equiv g^{\mu\nu} - p^\mu n^\nu - n^\mu p^\nu
\end{equation}
has been introduced since it serves as a projector, such that
\begin{align}
  \label{dimana:2GLeadTwist}
  A_1 &= \frac12 d_{\mu\nu}\, _2G^{\mu\nu} \,, \\
  \label{dimana:2GTwistFour}
  A_2 &= p_\mu p_\nu\, _2G^{\mu\nu} \,.
\end{align}
Now, since $\CLdim p=1$, $\CLdim d^{\mu\nu}=0$ and $\CLdim n=-1$, we conclude
with the help of \eqref{dimana:DimG} that $\CLdim A_2=\CLdim A_1+2$.  It
follows that in Eq.\,\eqref{dimana:Tensor2G1} the term proportional to $A_1$
gives the leading-twist contribution, while the term proportional $A_2$
contributes to twist-four, i.e.\ in \eqref{dimana:FacForm1a} the corresponding
expression is proportional to $\Lambda^2/Q^2$.
%
%
\subsection{Three Gluons}
The twist analysis of \eqref{dimana:FacForm1b} can be performed in complete
analogy. Exploring the tensor structure of $_3G^{\mu\sigma\nu}$ (defined in
Eq.\,\eqref{facres:G3matrixExp1}, below) one gets$^{\ref{epsFootn1}}$
\begin{equation} \label{dimana:Tensor3G1}
  _3G^{\mu\sigma\nu} = d^{\mu\nu}n^\sigma\, B^1_1 +
  d^{\mu\sigma}n^\nu\, B^2_1 + d^{\nu\sigma}n^\mu\, B^3_1 +
  n^\mu n^\nu n^\sigma B_2 \,.
\end{equation}
Here, we have made use of the fact that
\begin{equation} \label{dimana:VanG1}
    n_{\alpha_i}\,_3G^{\alpha_1\alpha_2\alpha_3}=0 \,.
\end{equation}
As mentioned before we postpone the derivation of some of the results we are
using here to the next section. Eq.\,\eqref{dimana:VanG1} follows since the
indices of $_3G^{\mu\sigma\nu}$ belong either to terms like $A^\alpha$ or like
$\omega_{\alpha'}^\alpha\partial^{\alpha'}$, where the tensor
$\omega_{\alpha'}^\alpha$ is introduced in \eqref{facres:omegaDef1}.  It
subtracts the longitudinal component of the momentum vector it is applied to
and $n_\alpha\omega_{\alpha'}^\alpha=0$. The various terms in
\eqref{dimana:Tensor3G1} can be projected out in the following way
\begin{align}
  \label{dimana:3GTwistFour}
  B_1^1 &= \frac12 d_{\mu\nu}p_\sigma\, _3G^{\mu\sigma\nu} \quad\text{etc.,} \\
  \label{dimana:3GTwistSix}
    B_2 &= p_\mu p_\nu p_\sigma\, _3G^{\mu\sigma\nu} \,,
\end{align}
and a dimensional analysis shows that the terms proportional $B_1^i$ contribute
to twist-four, while the term proportional $B_2$ contributes to twist-six.
%
%
\subsection{Four Gluons}
The tensor structure of $_4G^{\mu\sigma\lambda\nu}$ (see
Eqs.\,\eqref{facres:G41}-\eqref{facres:G43} below) is given by
\begin{equation} \label{dimana:Tensor4G1}
  _4G^{\mu\sigma\lambda\nu} = d^{\mu\nu}d^{\sigma\lambda} C_1^1 + \text{perm.}
  + d^{\mu\nu}n^\sigma n^\lambda C_2^1 + \text{perm.}
  + n^\mu n^\nu n^\sigma n^\lambda C_3 \,,
\end{equation}
with
\begin{align}
  \label{dimana:4GTwistFour}
  C_1^1 &= d_{\mu\nu\sigma\lambda}\, _4G^{\mu\sigma\lambda\nu}\quad\text{etc.,}
  \\
  \label{dimana:4GTwistSix}
  C_2^1 &= \frac12 d_{\mu\nu}p_\sigma p_\lambda\, _4G^{\mu\sigma\lambda\nu}
    \quad \text{etc.,} \\
  \label{dimana:4GTwistEight}
  C_3 &= p_\mu p_\nu p_\sigma p_\lambda\, _4G^{\mu\sigma\lambda\nu} \,,
\end{align}
where
\begin{equation}
  d_{\mu\nu\sigma\lambda}\equiv
  \frac18\left( 3d_{\mu\nu}d_{\sigma\lambda} - d_{\mu\sigma}d_{\nu\lambda} -
    d_{\mu\lambda}d_{\nu\sigma} \right) \,.
\end{equation}
In this case, the functions $C_1^i$ give the twist-four and the functions
$C_2^i$ the twist-six contributions, while $C_3$ belongs to twist-eight.

Terms with more than four $t$-channel gluons and/or derivatives can be analysed
in the same way. The term with lowest dimension of the corresponding five-gluon
contribution $_5G^{\mu\sigma\lambda\rho\nu}$ would be proportional
$d^{\mu\nu}d^{\sigma\lambda}n^\rho$ + (perm.) and, therefore, belongs to
twist-six.  Increasing the number of gluon-lines or derivatives also increases
the twist.  We can, therefore, justify that we have restricted ourselves to the
diagrams in Fig.\,\ref{FactDiags1} and to the terms in
Eqs.\,\eqref{fou:Rexp1}-\eqref{fou:Rexp3}.


%% file: facres.tx
\section{Results of the Factorization Procedure} \label{Sec:facres}
Before we present the results of the factorization procedure, let us summarize
the preceding results. Our computations are based on the diagrams in
Fig.\,\ref{FactDiags1}, where the expressions for the quark-loops $R$ are
expanded about the longitudinal components of the gluon-momenta
\eqref{fou:Rexp1}-\eqref{fou:Rexp3}. For the derivatives in \eqref{fou:Rexp1}
and \eqref{fou:Rexp2} we have computed some Ward identities in
section~\ref{Sec:Ward}, viz.\ \eqref{ward:TwoThreeRel1},
\eqref{ward:ThreeFourRel1}, \eqref{ward:ThreeFourRel2} and
\eqref{ward:TwoTwiceDiff1}. For the derivation of these relations we had to
explore the color structure of the quark-loop expressions,
\eqref{ward:twoGluonColStruk1}, \eqref{ward:threeGluonColStruk1} and
\eqref{ward:FourGcolorDecomp1}, which also led to an important relation for the
four-gluon quark-loop, \eqref{ward:VanishingRelFourQL1}, if one of the
gluon-momenta vanishes. In section~\ref{Sec:Ops} we have developed a method
which is useful in order to replace, within matrix elements, $A$-operators by
$F$-operators, and in section~\ref{Sec:DimAnalysis} we performed the
dimensional analysis of the expressions under consideration. This yielded a
scheme for projecting out contributions with definite twist.

As outlined before, it is our aim to write the gluonic twist-four contributions
to the inclusive deep-inelastic structure functions in an explicitly gauge
invariant form. In order to get these expressions it is, now, necessary to
reorder the terms which enter into the sum of the three diagrams in
Fig.\,\ref{FactDiags1} after expanding the quark loops about the longitudinal
components of the gluon-momenta as in
Eqs.\,\eqref{fou:Rexp1}-\eqref{fou:Rexp3}.  In particular, we write the sum of
these contributions as
\begin{equation} \label{facres:Hidef}
  \tilde H_1 + \tilde H_2 + \tilde H_3 =
  H_1 + H_2 + H_3 + \mathcal O \left( \frac1{Q^4} \right) \,.
\end{equation}
Here $H_1$ consists only of the zeroth-order term of the Taylor expansion of
the two-gluon quark-loop in Eq.\,\eqref{fou:Rexp1} inserted in
\eqref{fou:tildeH1}, while $H_2$ is the sum of the first-order two-gluon
quark-loop (from Eq.\,\eqref{fou:Rexp1}) and the zeroth-order three-gluon
quark-loop (from Eq.\,\eqref{fou:Rexp2}) inserted in Eqs.\,\eqref{fou:tildeH1}
and \eqref{fou:tildeH2}. $H_3$ consists of the sum of expressions
\eqref{fou:tildeH1} through \eqref{fou:tildeH3} where only the zeroth-order
term of the four-gluon quark-loop \eqref{fou:Rexp3}, the first-order
three-gluon quark-loop \eqref{fou:Rexp2} and the second-order two-gluon
quark-loop \eqref{fou:Rexp1} are taken into account. With these conventions the
steps needed to write $H_1$ and $H_2$ in an explicitly gauge invariant form are
straightforward, while the situation is more peculiar in the case of $H_3$. We
review the results in the next subsections.
%
%
\subsection{Two and Three Gluons} \label{Sec:facres:TwoThree}
Inserting the zeroth-order term in Eq.\,\eqref{fou:Rexp1} into
\eqref{fou:tildeH1}, we realize that the expression for the quark-loop depends
only on the longitudinal component of the four-momentum $k_1$, and we can
simplify the integrations over $k_1$ and $z$ if we introduce the expression
(cf.\,\eqref{intro:Sudakov1}) \cite{biblio:EFP}
\begin{equation}
  \int\!\!\CLrd^4 k_1 = \int\!\!\CLrd^4 k_1 \CLrd x_1
  \delta(x_1-k_1\cdot n)
\end{equation}
and make use of the relation
\begin{equation}
  \int\!\frac{\CLrd^4 k_1}{(2\pi)^4} \CLre^{\CLri k_1\cdot z}
    \delta(x_1-k_1\cdot n) =
  \int\!\frac{\CLrd\lambda}{2\pi} \CLre^{\CLri\lambda x_1}
    \delta^{(4)}(z-\lambda n) \,.
\end{equation}
The result is
\begin{equation}
  H_1 = \int\!\!\CLrd x_1 R_{\mu\nu}^{ab}(x_1p)
  \int\!\frac{\CLrd\lambda}{2\pi} \CLre^{\CLri\lambda x_1}
  \CLpBra\CLrT\left\{ A_b^\nu(0) A_a^\mu(\lambda n) \right\} \CLpKet \,.
\end{equation}
With the help of the results of section~\ref{Sec:Ops} we can substitute the
$A$-operators by $F$-operators, and with the help of
Eqs.\,\eqref{dimana:Tensor2G1} and \eqref{dimana:2GTwistFour} we can project
out the twist-four contribution:
\begin{equation} \label{facres:H1gaugeInv1}
  \left. H_1\right|_{\tau=4} = \int\!\! \CLrd x_1
  n^\mu n^\nu R_{\mu\nu}(x_1p) \,
  p_{\mu'}p_{\nu'}\, _2G^{\mu'\nu'}(x_1) \,,
\end{equation}
where
\begin{equation} \label{facres:G2matrixExp1}
  _2G^{\mu\nu}(x_1)= \frac{n_\alpha n_\beta}{x_1^2}
  \int\!\frac{\CLrd\lambda}{2\pi} \CLre^{\CLri\lambda x_1}
  \CLpBra\CLrT \left\{ \delta^{ab}F_b^{\nu\beta}(0)F_a^{\mu\alpha}(\lambda n)
  \right\} \CLpKet \,.
\end{equation}

In the case of $H_2$ we get the additional factor $(k_1-x_1p)^\sigma$ from the
first-order term in Eq.\,\eqref{fou:Rexp1}. This factor can be written as a
derivative applied to the exponential in \eqref{fou:GammaDef1a} if we introduce
the tensor \cite{biblio:EFP}
\begin{equation} \label{facres:omegaDef1}
  \omega_{\mu'}^\mu \equiv g_{\mu'}^\mu - p^\mu n_{\mu'} \,, \qquad
  \omega_{\mu'}^\mu k_1^{\mu'} = (k_1-x_1p)^\mu \,,
\end{equation}
which subtracts the longitudinal component of the momentum-vector it is applied
to. Now, using partial integration in order to apply this derivative to the
gluon operator and making use of Eq.\,\eqref{ward:TwoThreeRel1} we get
\begin{equation} \label{facres:H2withAs}
  H_2 = \int\!\!\CLrd x_1 \CLrd x_2 R_{\mu\sigma\nu}(x_1p,x_2p)
  \int\!\frac{\CLrd\lambda}{2\pi}\frac{\CLrd\eta}{2\pi}
  \CLre^{\CLri\lambda x_1}\CLre^{\CLri\eta(x_2-x_1)}
  \omega_{\sigma'}^\sigma
  \CLpBra\CLrT \left\{ A_c^\nu(0) \CLrD_{ac}^{\sigma'}(\eta n)
    A_a^\mu(\lambda n) \right\} \CLpKet \,,
\end{equation}
where the partial derivative within the covariant derivative is due to the
first-order term in the Taylor expansion \eqref{fou:Rexp1}, and the gluon
operator within $\CLrD_{ac}^{\sigma'}$ is due to the zeroth-order term in
\eqref{fou:Rexp2}.  For the derivation of \eqref{facres:H2withAs} we have also
made use of the fact that, within the axial gauge, the $\omega$-tensor applied
to a gluon-operator yields the same operator:
\begin{equation}
  \omega_{\mu'}^\mu A^{\mu'}_a = A^\mu_a \,.
\end{equation}
Projecting out the twist-four contribution with the help of
Eqs.\,\eqref{dimana:Tensor3G1} and \eqref{dimana:3GTwistFour} and applying the
method developed in section~\ref{Sec:Ops} in order to substitute the
$A$-operators by $F$-operators we end up with
\begin{equation} \label{facres:H2gaugeInv1}
  \left. H_2\right|_{\tau=4} = \int\!\! \CLrd x_1\CLrd x_2
  \left\{ d^{\mu\nu}n^\sigma R_{\mu\sigma\nu}(x_1p,x_2p) \,
    \frac12 d_{\mu'\nu'}p_{\sigma'} \,
    _3G^{\mu'\sigma'\nu'}(x_1,x_2) + \text{perm.} \right\} \,,
\end{equation}
where
\begin{equation} \label{facres:G3matrixExp1}
  _3G^{\mu\sigma\nu}(x_1,x_2) = \frac{n_\alpha n_\beta}{x_1\,x_2}
  \int\! \frac{\CLrd\lambda}{2\pi}\frac{\CLrd\eta}{2\pi}
  \CLre^{\CLri\lambda x_1}\CLre^{\CLri\eta(x_2-x_1)}\omega_{\sigma'}^\sigma
  \CLpBra\CLrT \left\{ F_c^{\nu\beta}(0) 
    \CLrD_{ac}^{\sigma'}(\eta n)
    F_a^{\mu\alpha}(\lambda n) \right\} \CLpKet \,.
\end{equation}
For the permutation in \eqref{facres:H2gaugeInv1} one has to keep
Eqs.\,\eqref{dimana:Tensor3G1} and \eqref{dimana:3GTwistFour} in mind, i.e.\ 
indices of the projector which is applied to $R_{\mu\sigma\nu}$ and of the
projector which is applied to $_3G^{\mu'\sigma'\nu'}$ have to be permutated
simultaneously. Similar to $H_1$, \eqref{facres:H1gaugeInv1}, $H_2$ contains
only contributions of the two-gluon operator.

We expect that Eqs.\,\eqref{facres:H1gaugeInv1} and \eqref{facres:H2gaugeInv1}
are of more general validity, since the Ward identities we have used are valid
to all orders and our ans\"atze for the color structure of $R_{\mu\nu}^{ab}$
and $R_{\mu\lambda\nu}^{abc}$, \eqref{ward:twoGluonColStruk1} and
\eqref{ward:threeGluonColStruk1}, are of the most general nature.
%
%
\subsection{Four Gluons}
\subsubsection{Basics} \label{Sec:facres:FGBasics}
The situation is more complicated in the case of $H_3$. Putting together all
terms which enter into $H_3$ and making use of
Eqs.\,\eqref{ward:ThreeFourRel1}, \eqref{ward:ThreeFourRel2} and
\eqref{ward:TwoTwiceDiff1} we get
\begin{align} \label{facres:H3firstRep1}
  H_3 = \int\!\!\CLrd x_1 \CLrd x_2 \CLrd x_3 \Bigl\{
    & R_{\mu\sigma\lambda\nu}^{abcd}(x_1p,x_2p,x_3p)\,
      _4G_{abcd}^{\mu\sigma\lambda\nu}(x_1,x_2,x_3) + \notag\\
    +\, & \frac\CLri2 g f^{acd} R_{\mu\sigma\lambda\nu}^1(x_1p,x_2p,x_3p)\,
      _3G_{acd,1}^{\mu\sigma\lambda\nu}(x_1,x_3)\delta(x_1-x_2) + \notag\\
    +\, & \frac\CLri2 g f^{abd} R_{\mu\sigma\lambda\nu}^1(x_1p,x_2p,x_3p)\,
      _3G_{abd,2}^{\mu\sigma\lambda\nu}(x_1,x_2)\delta(x_2-x_3) + \\
    +\, &\frac14 \delta^{ad}
      \bigl[ R_{\mu\sigma\lambda\nu}^1(x_1p,x_2p,x_3p)+ \notag\\
        & \qquad\qquad
        + R_{\mu\sigma\lambda\nu}^3(x_1p,x_2p,x_3p) \bigr]\,
      _2G_{ad}^{\mu\sigma\lambda\nu}(x_1)
      \delta(x_1-x_2)\delta(x_2-x_3) \Bigr\} \,, \notag
\end{align}
where
\begin{align}
  \label{facres:4GfirstDef1}
  _4G_{abcd}^{\mu\sigma\lambda\nu}(x_1,x_2,x_3) &=
    \int\!\frac{\CLrd\lambda}{2\pi}\frac{\CLrd\eta}{2\pi}\frac{\CLrd\tau}{2\pi}
    \CLre^{\CLri\lambda x_1}\CLre^{\CLri\eta(x_2-x_1)}
    \CLre^{\CLri\tau(x_3-x_2)}
    \CLpBra\CLrT\left\{ A_d^\nu(0) A_c^\lambda(\tau n) A_b^\sigma(\eta n)
      A_a^\mu(\lambda n) \right\} \CLpKet \,, \\
  \label{facres:3GfirstDef1}
  _3G_{acd,1}^{\mu\sigma\lambda\nu}(x_1,x_3) &=
    \omega_{\sigma'}^\sigma \int\! \frac{\CLrd\lambda}{2\pi}
    \frac{\CLrd\tau}{2\pi}  \CLre^{\CLri\lambda x_1} \CLre^{\CLri\tau(x_3-x_2)}
    \CLpBra\CLrT\left\{ A_d^\nu(0) A_c^\lambda(\tau n) \CLri \partial^{\sigma'}
      A_a^\mu(\lambda n) \right\} \CLpKet \,, \\
  \label{facres:3GfirstDef2}
  _3G_{abd,2}^{\mu\sigma\lambda\nu}(x_1,x_2) &=
    \omega_{\lambda'}^\lambda \int\! \frac{\CLrd\lambda}{2\pi}
    \frac{\CLrd\eta}{2\pi} \CLre^{\CLri\lambda x_1}\CLre^{\CLri\eta(x_2-x_1)}
    \CLpBra\CLrT\left\{ A_d^\nu(0) \CLri\partial^{\lambda'} [A_b^\sigma(\eta n)
      A_a^\mu(\lambda n)] \right\} \CLpKet \,, \\
  \label{facres:2GfirstDef1}
  _2G_{ad}^{\mu\sigma\lambda\nu}(x_1) &=
    \omega_{\sigma'}^\sigma \omega_{\lambda'}^\lambda \int\!
    \frac{\CLrd\lambda}{2\pi} \CLre^{\CLri\lambda x_1}
    \CLpBra\CLrT\left\{ A_d^\nu(0) \CLri \partial^{\lambda'} \CLri
      \partial^{\sigma'} A_a^\mu(\lambda n) \right\} \CLpKet \,.
\end{align}
The derivatives in Eqs.\,\eqref{facres:3GfirstDef1}-\eqref{facres:2GfirstDef1}
have to be understood as $\partial^{\sigma'}A(\lambda n)\equiv\partial
A(z)/\partial z_{\sigma'}| _{z=\lambda n}$, etc.

Due to our definition \eqref{ward:FourGcolorDecomp1}, we can split up
Eq.\,\eqref{facres:H3firstRep1} into three parts
\begin{equation} \label{facres:H3decomp1}
  H_3 = H_3^1 + H_3^2 + H_3^3 \,,
\end{equation}
where $H_3^i$ corresponds to those terms in \eqref{facres:H3firstRep1} which
contain the function $R_{\mu\sigma\lambda\nu}^i$. Since we intend to write the
expression for $H_3$ in a form which is explicitly gauge invariant (similar to
\eqref{facres:H1gaugeInv1} and \eqref{facres:H2gaugeInv1}) we have to modify it
in such a way that partial derivatives occur only within covariant derivatives.
This can most easily be done with the help of the following relation:
\begin{equation} \label{facres:partDerivSubst1}
  f^{acd}A_c^\lambda(\tau n)\partial^\sigma =
  f^{a'cd}A_c^\lambda(\tau n)\CLrD_{a'a}^\sigma(\eta n) +
  g f^{aba'}f^{a'cd} A_c^\lambda(\tau n) A_b^\sigma(\eta n) \,.
\end{equation}
Inserting \eqref{facres:partDerivSubst1} into \eqref{facres:H3firstRep1} we
realize that the results for the functions $H_3^i$ consist of terms with two-,
three- and four-gluon operators. In the spin-averaged case the three-gluon
operator does not contribute. The decomposition \eqref{facres:H3decomp1} is not
unique, since the Ward identities could be expressed with the help of different
combinations of the functions $R_{\mu\sigma\lambda\nu}^i$ in
Eqs.\,\eqref{ward:ThreeFourRel1}, \eqref{ward:ThreeFourRel2} and
\eqref{ward:TwoTwiceDiff1}. We could, therefore, also write
Eq.\,\eqref{facres:H3firstRep1} with different combinations of these functions.
By exploring symmetry relations for
$R_{\mu\sigma\lambda\nu}^{abcd}(k_1,k_2,k_3)$ we will show how to derive
expressions for $H_3$ which involve only one of the functions $R^i$ and contain
no three-gluon operator.
%
%
\subsubsection{Symmetry Relations of the Four-Gluon Quark-Loop Expression}
\label{Sec:facres:FGSymmetry}
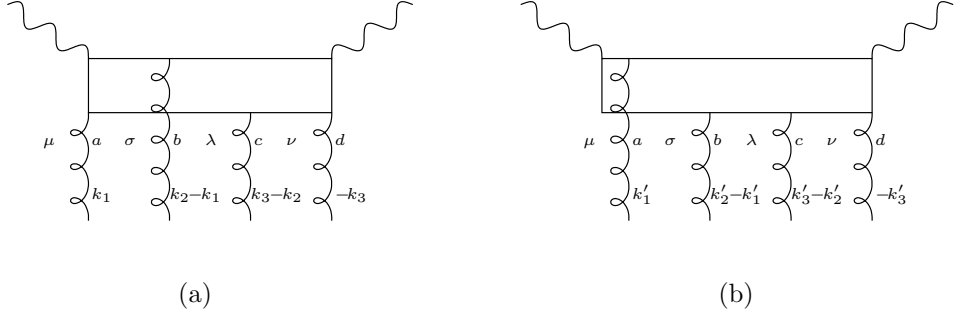
\begin{figure}
  \begin{center}
    \input{symexamp1.pstex_t}
    \caption{Two sample diagrams which enter into
      $R_{\mu\sigma\lambda\nu}^{abcd}$. They are identical if one sets
      $k_1=k_2'-k_1'$, $k_2=k_2'$, $k_3=k_3'$ and exchanges
      $\mu\leftrightarrow\sigma$ and $a\leftrightarrow b$.}
    \label{facres:SymFig1}
  \end{center}
\end{figure}

The expression for the four-gluon quark-loop,
$R_{\mu\sigma\lambda\nu}^{abcd}(k_1,k_2,k_3)$, has to be completely symmetrical
under the simultaneous exchange of gluon-momenta, Lorentz- and color indices.
This can be seen in two ways. First, within the path-integral formulation we
have defined $R_{\mu\sigma\lambda\nu}^{abcd}(k_1,k_2,k_3)$ in
Eq.\,\eqref{ward:PertQLdef3}. Since the gluon is a boson the derivatives in
\eqref{ward:PertQLdef3} commute and the proclaimed symmetry follows
immediately. Secondly, it can be seen diagrammatically. Clearly,
$R_{\mu\sigma\lambda\nu}^{abcd}(k_1,k_2,k_3)$ corresponds to the expression
which is obtained if one takes into account all diagrams up to a specified
order, i.e.\ in leading order one has to consider all permutations for the
coupling of four $t$-channel gluons to quarks and anti-quarks.  Consider, for
example, the diagram in Fig.\,\ref{facres:SymFig1}a. If we set $k_1=k_2'-k_1'$,
$k_2=k_2'$, $k_3=k_3'$ and exchange $\mu\leftrightarrow\sigma$ and
$a\leftrightarrow b$ we get an analytic expression which is the same as the one
we get from the diagram in Fig.\,\ref{facres:SymFig1}b. Since such a relation
can be found for each diagram that enters into $R_{\mu\sigma\lambda\nu}^{abcd}$
we have in general
\begin{equation}
  R_{\mu\sigma\lambda\nu}^{abcd}(k_1,k_2,k_3) =
  R_{\sigma\mu\lambda\nu}^{bacd}(k_2-k_1,k_2,k_3) \,.
\end{equation}
Next, with the help of the color structure \eqref{ward:FourGcolorDecomp1} of
$R_{\mu\sigma\lambda\nu}^{abcd}$ we can find relations between the three
functions $R_{\mu\sigma\lambda\nu}^i$ if we realize that for $a\leftrightarrow
b$: $d^{abcd}\to d^{abdc}$, $d^{abdc}\to d^{abcd}$ but $d^{acbd}\to d^{acbd}$,
as follows from the definition of $d^{abcd}$, \eqref{ward:dabcdDef1}. With
these relations we get
\begin{align}
  R_{\mu\sigma\lambda\nu}^1(k_1,k_2,k_3) &=
    R_{\sigma\mu\lambda\nu}^2(k_2-k_1,k_2,k_3) \,, \\
  R_{\mu\sigma\lambda\nu}^2(k_1,k_2,k_3) &=
    R_{\sigma\mu\lambda\nu}^1(k_2-k_1,k_2,k_3) \,, \\
  R_{\mu\sigma\lambda\nu}^3(k_1,k_2,k_3) &=
    R_{\sigma\mu\lambda\nu}^3(k_2-k_1,k_2,k_3) \,.
\end{align}
Exchanging other gluon lines one gets a large number of relations. Those which
are useful in the subsequent derivations and a few additional ones are given in
Table~\ref{facres:SymTab1}.
%
%
\begin{table}
  \begin{center}
    \begin{small}
      \begin{tabular}{r|l|l|l}
        & $R_{\mu\sigma\lambda\nu}^1(k_1,k_2,k_3)=$ & 
          $R_{\mu\sigma\lambda\nu}^2(k_1,k_2,k_3)=$ & 
          $R_{\mu\sigma\lambda\nu}^3(k_1,k_2,k_3)=$ \\[3pt]
        \hline
        && \\[-7pt]
        1 & $R_{\sigma\mu\lambda\nu}^2(k_2-k_1,k_2,k_3)$ &
          $R_{\sigma\mu\lambda\nu}^1(k_2-k_1,k_2,k_3)$ &
          $R_{\sigma\mu\lambda\nu}^3(k_2-k_1,k_2,k_3)$ \\[3pt]
        2 & $R_{\mu\sigma\nu\lambda}^2(k_1,k_2,k_2-k_3)$ &
          $R_{\mu\sigma\nu\lambda}^1(k_1,k_2,k_2-k_3)$ &
          $R_{\mu\sigma\nu\lambda}^3(k_1,k_2,k_2-k_3)$ \\[3pt]
        3 & $R_{\mu\lambda\sigma\nu}^3(k_1,k_1-k_2+k_3,k_3)$ &
          $R_{\mu\lambda\sigma\nu}^2(k_1,k_1-k_2+k_3,k_3)$ &
          $R_{\mu\lambda\sigma\nu}^1(k_1,k_1-k_2+k_3,k_3)$ \\[3pt]
        4 & $R_{\sigma\lambda\mu\nu}^2(k_2-k_1,k_3-k_1,k_3)$ &
          $R_{\sigma\lambda\mu\nu}^3(k_2-k_1,k_3-k_1,k_3)$ &
          $R_{\sigma\lambda\mu\nu}^1(k_2-k_1,k_3-k_1,k_3)$ \\[3pt]
        5 & $R_{\lambda\mu\sigma\nu}^3(k_3-k_2,k_1-k_2+k_3,k_3)$ &
          $R_{\lambda\mu\sigma\nu}^1(k_3-k_2,k_1-k_2+k_3,k_3)$ &
          $R_{\lambda\mu\sigma\nu}^2(k_3-k_2,k_1-k_2+k_3,k_3)$ \\[3pt]
        6 & $R_{\lambda\sigma\mu\nu}^1(k_3-k_2,k_3-k_1,k_3)$ &
          $R_{\lambda\sigma\mu\nu}^3(k_3-k_2,k_3-k_1,k_3)$ &
          $R_{\lambda\sigma\mu\nu}^2(k_3-k_2,k_3-k_1,k_3)$ \\[3pt]
        7 & $R_{\mu\nu\lambda\sigma}^1(k_1,k_1-k_3,k_1-k_2)$ &
          $R_{\mu\nu\lambda\sigma}^3(k_1,k_1-k_3,k_1-k_2)$ &
          $R_{\mu\nu\lambda\sigma}^2(k_1,k_1-k_3,k_1-k_2)$ \\[3pt]
      \end{tabular}
    \end{small}
    \caption{Making use of symmetry relations for
      $R_{\mu\sigma\lambda\nu}^{abcd}(k_1,k_2,k_3)$ one finds relations
      between the three functions $R_{\mu\sigma\lambda\nu}^i(k_1,k_2,k_3)$.}
    \label{facres:SymTab1}
  \end{center}
\end{table}
%
%
%
%
\subsubsection{Gauge Invariant Expressions} \label{Sec:facres:FGResults}
With the help of the relations in Table~\ref{facres:SymTab1} we can, now, write
Eq.\,\eqref{facres:H3firstRep1} as an explicitly gauge invariant expression.
For example, using the relation in row one, column two of
Table~\ref{facres:SymTab1} we can, in the first line of
\eqref{facres:H3firstRep1}, replace the function $R_{\mu\sigma\lambda\nu}^2$ by
$R_{\sigma\mu\lambda\nu}^1$ and after an appropriate transformation of
integration variables ($x_1\to x_2-x_1$, $\lambda\leftrightarrow\eta$) and an
exchange of Lorentz- and color indices $R_{\mu\sigma\lambda\nu}^1$ can be
factored out.  Similarly the terms involving $R_{\mu\sigma\lambda\nu}^3$ in
\eqref{facres:H3firstRep1} can be brought into the desired form with the help
of the expression in row three, column three of Table~\ref{facres:SymTab1}.  We
end up with
\begin{multline} \label{facres:H3firstExpr1}
  H_3 = \int\!\!\CLrd x_1\CLrd x_2\CLrd x_3
    \int\!\frac{\CLrd\lambda}{2\pi}\frac{\CLrd\eta}{2\pi}\frac{\CLrd\tau}{2\pi}
    \CLre^{\CLri\lambda x_1}\CLre^{\CLri\eta(x_2-x_1)}
    \CLre^{\CLri\tau(x_3-x_2)}
    R_{\mu\sigma\lambda\nu}^1(x_1p,x_2p,x_3p)
    \omega_{\sigma'}^\sigma \omega_{\lambda'}^\lambda \times \\
  \times\CLpBra\CLrT\bigl\{ A_d^\nu(0)\bigl[
    g^2l_1^{abcd} A_c^{\lambda'}(\tau n) A_b^{\sigma'}(\eta n)-
    \frac12 \CLrD_{ac'}^{\lambda'}(\tau n) \CLrD_{c'd}^{\sigma'}(\eta n)
    \bigr] A_a^\mu(\lambda n) \bigr\} \CLpKet \,,
\end{multline}
where
\begin{align} \label{facres:ColFacl1}
  l_1^{abcd} &= f^{adc'}f^{c'bc} +
    \frac14 \left( \delta^{ab}\delta^{cd} + \delta^{ac}\delta^{bd} +
      \delta^{ad}\delta^{bc} \right) = \\
  &= \frac12 f^{acc'}f^{c'bd} - \frac32 d^{acc'}d^{c'bd} +
    \frac12 \left( \delta^{ab}\delta^{cd} - \delta^{ac}\delta^{bd} +
      \delta^{ad}\delta^{bc} \right) +
    \frac14 \left( \delta^{ab}\delta^{cd} + \delta^{ac}\delta^{bd} +
      \delta^{ad}\delta^{bc} \right) = \notag\\
  &= \frac12 f^{acc'}f^{c'bd} + 3d^{abcd} \notag\,.
\end{align}
From Eq.\,\eqref{facres:H3firstRep1} we realize that there is a term
proportional $R_{\mu\sigma\lambda\nu}^{abcd}$ in the first line.  Due to
\eqref{ward:FourGcolorDecomp1} this yields terms with color factors $d^{abcd}$
and corresponding permutations of color indices. For the derivation of
\eqref{facres:H3firstExpr1} one has to add and subtract terms which involve
four $A$-operators in order to get covariant derivatives. Combining the
subtracted expressions with those in the first line of
Eq.\,\eqref{facres:H3firstRep1} the resulting color factor is no longer
proportional to $d^{abcd}$. Instead one gets the term in
Eq.\,\eqref{facres:ColFacl1}.

Again, with the help of \eqref{dimana:Tensor4G1} and \eqref{dimana:4GTwistFour}
we can project out the twist-four contributions and with the help of the
results in section~\ref{Sec:Ops} we can substitute the $A$-operators by
$F$-operators:
\begin{equation}
  \label{facres:H3gaugeInv1}
  \left. H_3\right|_{\tau=4} = \int\!\!\CLrd x_1\CLrd x_2\CLrd x_3
  \left\{ d^{\mu\nu}d^{\sigma\lambda} R_{\mu\sigma\lambda\nu}^1(x_1p,x_2p,x_3p)
    \, d_{\mu'\nu'\sigma'\lambda'}\,
    _4G_1^{\mu'\sigma'\lambda'\nu'}(x_1,x_2,x_3)
    + \text{perm.} \right\} \,,
\end{equation}
where
\begin{align}
  \label{facres:G41}
  _4&G_1^{\mu\sigma\lambda\nu}(x_1,x_2,x_3) =
    \int\!\frac{\CLrd\lambda}{2\pi}\frac{\CLrd\eta}{2\pi}\frac{\CLrd\tau}{2\pi}
    \CLre^{\CLri\lambda x_1}\CLre^{\CLri\eta(x_2-x_1)}
    \CLre^{\CLri\tau(x_3-x_2)} \times \\ & \,\,\times \Biggl[
  \frac{-1}{x_1x_3(x_2-x_1)(x_3-x_2)} 
    n_\alpha n_\beta n_\gamma n_\delta \CLpBra\CLrT\bigl\{
    g^2l_1^{abcd} F_d^{\nu\delta}(0) F_c^{\lambda\gamma}(\tau n)
    F_b^{\sigma\beta}(\eta n) F_a^{\mu\alpha}(\lambda n)\bigr\}
    \CLpKet \notag \\
  & \,\,\qquad-\frac12\frac{1}{x_1x_3} n_\alpha n_\beta \CLpBra\CLrT\bigl\{
    F_d^{\nu\beta}(0)\omega_{\lambda'}^\lambda
    \CLrD_{ac}^{\lambda'}(\tau n) \omega_{\sigma'}^\sigma
    \CLrD_{cd}^{\sigma'}(\eta n) F_a^{\mu\alpha}(\lambda n)
    \bigr\} \CLpKet \Biggr] \notag \,,
\end{align}
and the permutations in \eqref{facres:H3gaugeInv1} have to be understood in the
same sense as in \eqref{facres:H2gaugeInv1} (cf.\ \eqref{dimana:Tensor4G1},
\eqref{dimana:4GTwistFour}). For $H_3$ we get contributions of the two-gluon
operator as well as of the four-gluon operator.

The representation of $H_3$ in a gauge invariant form as in
Eq.\,\eqref{facres:H3gaugeInv1} is not unique. For example, we can derive
expressions where the quark-loop is represented by $R^2$ or $R^3$ instead of
$R^1$. These expressions can be obtained in two ways. One can either start with
Eq.\,\eqref{facres:H3firstRep1}, make use of the relations in
Table~\ref{facres:SymTab1} and apply the same methods which led to
\eqref{facres:H3gaugeInv1}. Or one starts with \eqref{facres:H3gaugeInv1} and
rewrites $R^1$ in terms of $R^2$ or $R^3$ with the help of the expressions in
Table~\ref{facres:SymTab1}. The results can be summarized as follows. If we
exchange $R_{\mu\sigma\lambda\nu}^1(x_1p,x_2p,x_3p)$ by
$R_{\mu\sigma\lambda\nu}^2(x_1p,x_2p,x_3p)$ or
$R_{\mu\sigma\lambda\nu}^3(x_1p,x_2p,x_3p)$ in \eqref{facres:H3gaugeInv1} we
also have to exchange $_4G_1^{\mu\sigma\lambda\nu}(x_1,x_2,x_3)$ by
$_4G_2^{\mu\sigma\lambda\nu}(x_1,x_2,x_3)$ or
$_4G_3^{\mu\sigma\lambda\nu}(x_1,x_2,x_3)$, resp. The latter functions can be
written as
\begin{align} \label{facres:G42}
  _4&G_2^{\mu\sigma\lambda\nu}(x_1,x_2,x_3) =
    \int\!\frac{\CLrd\lambda}{2\pi}\frac{\CLrd\eta}{2\pi}\frac{\CLrd\tau}{2\pi}
    \CLre^{\CLri\lambda x_1}\CLre^{\CLri\eta(x_2-x_1)}
    \CLre^{\CLri\tau(x_3-x_2)} \times \\ & \,\,\times \Biggl[
  \frac{-1}{x_1x_3(x_2-x_1)(x_3-x_2)} 
    n_\alpha n_\beta n_\gamma n_\delta \CLpBra\CLrT\bigl\{
    g^2l_2^{abcd} F_d^{\nu\delta}(0) F_c^{\lambda\gamma}(\tau n)
    F_b^{\sigma\beta}(\eta n) F_a^{\mu\alpha}(\lambda n)\bigr\}
    \CLpKet \notag \\
  & \,\,\qquad-\frac12\frac{1}{(x_2-x_1)x_3} n_\alpha n_\beta
    \CLpBra\CLrT\bigl\{
    F_d^{\nu\beta}(0)\omega_{\lambda'}^\lambda
    \CLrD_{bc}^{\lambda'}(\tau n) \omega_{\mu'}^\mu
    \CLrD_{cd}^{\mu'}(\lambda n) F_b^{\sigma\alpha}(\eta n)
    \bigr\} \CLpKet \Biggr] \notag
\end{align}
and
\begin{align} \label{facres:G43}
  _4&G_3^{\mu\sigma\lambda\nu}(x_1,x_2,x_3) =
    \int\!\frac{\CLrd\lambda}{2\pi}\frac{\CLrd\eta}{2\pi}\frac{\CLrd\tau}{2\pi}
    \CLre^{\CLri\lambda x_1}\CLre^{\CLri\eta(x_2-x_1)}
    \CLre^{\CLri\tau(x_3-x_2)} \times \\ & \,\,\times \Biggl[
  \frac{-1}{x_1x_3(x_2-x_1)(x_3-x_2)} 
    n_\alpha n_\beta n_\gamma n_\delta \CLpBra\CLrT\bigl\{
    g^2l_3^{abcd} F_d^{\nu\delta}(0) F_c^{\lambda\gamma}(\tau n)
    F_b^{\sigma\beta}(\eta n) F_a^{\mu\alpha}(\lambda n)\bigr\}
    \CLpKet \notag \\
  & \,\,\qquad-\frac12\frac{1}{x_1x_3} n_\alpha n_\beta \CLpBra\CLrT\bigl\{
    F_d^{\nu\beta}(0)\omega_{\sigma'}^\sigma
    \CLrD_{ac}^{\sigma'}(\eta n) \omega_{\lambda'}^\lambda
    \CLrD_{cd}^{\lambda'}(\tau n) F_a^{\mu\alpha}(\lambda n)
    \bigr\} \CLpKet \Biggr] \,, \notag
\end{align}
where
\begin{align} \label{facres:ColFacl2}
  l_2^{abcd} &= f^{acc'}f^{c'bd} +
    \frac14 \left( \delta^{ab}\delta^{cd} + \delta^{ac}\delta^{bd} +
      \delta^{ad}\delta^{bc} \right) = \\
  &= \frac12 f^{adc'}f^{c'bc} - \frac32 d^{adc'}d^{c'bc} +
    \frac12 \left( \delta^{ab}\delta^{cd} + \delta^{ac}\delta^{bd} -
      \delta^{ad}\delta^{bc} \right) +
    \frac14 \left( \delta^{ab}\delta^{cd} + \delta^{ac}\delta^{bd} +
      \delta^{ad}\delta^{bc} \right) \notag
\intertext{and} \label{facres:ColFacl3}
  l_3^{abcd} &= -f^{adc'}f^{c'bc} +
    \frac14 \left( \delta^{ab}\delta^{cd} + \delta^{ac}\delta^{bd} +
      \delta^{ad}\delta^{bc} \right) = \\
  &= \frac12 f^{abc'}f^{c'cd} - \frac32 d^{abc'}d^{c'cd} +
    \frac12 \left( -\delta^{ab}\delta^{cd} + \delta^{ac}\delta^{bd} +
      \delta^{ad}\delta^{bc} \right) +
    \frac14 \left( \delta^{ab}\delta^{cd} + \delta^{ac}\delta^{bd} +
      \delta^{ad}\delta^{bc} \right) \,. \notag
\end{align}
The color factors $l_i^{abcd}$ are related by permutations of color indices:
\begin{equation}
  l_1^{abcd} = l_2^{abdc} = l_3^{acbd} \,.
\end{equation}
Note the different Lorentz- and color structure and the different dependence on
integration variables of the two-gluon operators in \eqref{facres:G41},
\eqref{facres:G42} and \eqref{facres:G43}.
%
%
\subsubsection{Color Structure} \label{Sec:facres:Col}
As we have seen, $H_3$ contains contributions of the two- and four-gluon
operators. In this subsection we will explore the color factors $l_i^{abcd}$,
which enter into the contribution of the four-gluon operator and are given by
Eqs.\,\eqref{facres:ColFacl1}, \eqref{facres:ColFacl2} and
\eqref{facres:ColFacl3} depending on which representation of $H_3$ we are
considering. If we have a look at Table~\ref{facres:SymTab1}, we realize that
there are also relations which transform one of the functions $R^i$ into the
same function $R^i$ and only the dependence on the momenta and the
Lorentz-indices changes.  For example, if we start with the \textit{four-gluon
  operator} part of \eqref{facres:H3gaugeInv1}, make use of the relation in
line seven, row one of Table~\ref{facres:SymTab1}, exchange the integration
variables such that $x_3\to x_1-x_2$, $x_2\to x_1-x_3$,
$\lambda\to\lambda-\eta$, $\eta\to-\eta$ and $\tau\to\tau-\eta$ and make use of
Lorentz-invariance we get an expression which has exactly the same form for the
\textit{four-gluon operator} part as in Eq.\,\eqref{facres:G41} but the color
factor has to be exchanged by $l_1^{adcb}$, i.e.\ indices $b$ and $d$ are
exchanged. Now, the color factor $l_1^{abcd}$ in \eqref{facres:ColFacl1} has
been written in three ways. There is the compact representation in the first
line of \eqref{facres:ColFacl1}, and in the second line it has been split up
into symmetric and anti-symmetric parts with respect of exchange of color
indices $a$ and $c$ or $b$ and $d$. The anti-symmetric part is given by the
expression proportional $f^{acc'}f^{c'bd}$.  Since the expression we got after
use of the relation in Table~\ref{facres:SymTab1} must be the same as in
\eqref{facres:H3gaugeInv1}, we conclude that the term proportional
$f^{acc'}f^{c'bd}$ in \eqref{facres:ColFacl1} within \eqref{facres:H3gaugeInv1}
must give a vanishing contribution. In other words, the color factor that
survives has the form
\begin{equation} \label{facres:SurvColFac1}
  \tilde l_1^{abcd} = 3d^{abcd} \,.
\end{equation}

We can repeat this argument with those expressions in which $H_3$ is expressed
with the help of $R^2$ or $R^3$, \eqref{facres:G42}, \eqref{facres:G43}.
Again, the surviving color tensors have the form
\begin{align}
  \label{facres:SurvColFac2}
  \tilde l_2^{abcd} &= 3d^{abdc} \,, \\
  \label{facres:SurvColFac3}
  \tilde l_3^{abcd} &= 3d^{acbd} \,.
\end{align}
%
%


%% file: symexamp1.pstex_t
\begin{picture}(0,0)%
\epsfig{file=symexamp1.pstex}%
\end{picture}%
\setlength{\unitlength}{3947sp}%
\begingroup\makeatletter\ifx\SetFigFont\undefined%
\gdef\SetFigFont#1#2#3#4#5{%
  \reset@font\fontsize{#1}{#2pt}%
  \fontfamily{#3}\fontseries{#4}\fontshape{#5}%
  \selectfont}%
\fi\endgroup%
\begin{picture}(5964,1925)(214,-1494)
\put(1312,-1467){\makebox(0,0)[lb]{\smash{\SetFigFont{9}{10.8}{\rmdefault}{\mddefault}{\updefault}{\small (a)}}}}
\put(4706,-1467){\makebox(0,0)[lb]{\smash{\SetFigFont{9}{10.8}{\rmdefault}{\mddefault}{\updefault}{\small (b)}}}}
\put(464,-483){\makebox(0,0)[lb]{\smash{\SetFigFont{9}{10.8}{\rmdefault}{\mddefault}{\updefault}{\tiny $\mu$}}}}
\put(1991,-483){\makebox(0,0)[lb]{\smash{\SetFigFont{9}{10.8}{\rmdefault}{\mddefault}{\updefault}{\tiny $\nu$}}}}
\put(2296,-483){\makebox(0,0)[lb]{\smash{\SetFigFont{9}{10.8}{\rmdefault}{\mddefault}{\updefault}{\tiny $d$}}}}
\put(1278,-483){\makebox(0,0)[lb]{\smash{\SetFigFont{9}{10.8}{\rmdefault}{\mddefault}{\updefault}{\tiny $b$}}}}
\put(1787,-483){\makebox(0,0)[lb]{\smash{\SetFigFont{9}{10.8}{\rmdefault}{\mddefault}{\updefault}{\tiny $c$}}}}
\put(769,-483){\makebox(0,0)[lb]{\smash{\SetFigFont{9}{10.8}{\rmdefault}{\mddefault}{\updefault}{\tiny $a$}}}}
\put(973,-483){\makebox(0,0)[lb]{\smash{\SetFigFont{9}{10.8}{\rmdefault}{\mddefault}{\updefault}{\tiny $\sigma$}}}}
\put(1482,-483){\makebox(0,0)[lb]{\smash{\SetFigFont{9}{10.8}{\rmdefault}{\mddefault}{\updefault}{\tiny $\lambda$}}}}
\put(5691,-483){\makebox(0,0)[lb]{\smash{\SetFigFont{9}{10.8}{\rmdefault}{\mddefault}{\updefault}{\tiny $d$}}}}
\put(5182,-483){\makebox(0,0)[lb]{\smash{\SetFigFont{9}{10.8}{\rmdefault}{\mddefault}{\updefault}{\tiny $c$}}}}
\put(4672,-483){\makebox(0,0)[lb]{\smash{\SetFigFont{9}{10.8}{\rmdefault}{\mddefault}{\updefault}{\tiny $b$}}}}
\put(4163,-483){\makebox(0,0)[lb]{\smash{\SetFigFont{9}{10.8}{\rmdefault}{\mddefault}{\updefault}{\tiny $a$}}}}
\put(5385,-483){\makebox(0,0)[lb]{\smash{\SetFigFont{9}{10.8}{\rmdefault}{\mddefault}{\updefault}{\tiny $\nu$}}}}
\put(4876,-483){\makebox(0,0)[lb]{\smash{\SetFigFont{9}{10.8}{\rmdefault}{\mddefault}{\updefault}{\tiny $\lambda$}}}}
\put(4367,-483){\makebox(0,0)[lb]{\smash{\SetFigFont{9}{10.8}{\rmdefault}{\mddefault}{\updefault}{\tiny $\sigma$}}}}
\put(3858,-483){\makebox(0,0)[lb]{\smash{\SetFigFont{9}{10.8}{\rmdefault}{\mddefault}{\updefault}{\tiny $\mu$}}}}
\put(2296,-822){\makebox(0,0)[lb]{\smash{\SetFigFont{9}{10.8}{\rmdefault}{\mddefault}{\updefault}{\tiny $-\!k_3$}}}}
\put(1787,-822){\makebox(0,0)[lb]{\smash{\SetFigFont{9}{10.8}{\rmdefault}{\mddefault}{\updefault}{\tiny $\!k_3\!\!-\!\!k_2$}}}}
\put(1278,-822){\makebox(0,0)[lb]{\smash{\SetFigFont{9}{10.8}{\rmdefault}{\mddefault}{\updefault}{\tiny $\!k_2\!\!-\!\!k_1$}}}}
\put(769,-822){\makebox(0,0)[lb]{\smash{\SetFigFont{9}{10.8}{\rmdefault}{\mddefault}{\updefault}{\tiny $k_1$}}}}
\put(5691,-822){\makebox(0,0)[lb]{\smash{\SetFigFont{9}{10.8}{\rmdefault}{\mddefault}{\updefault}{\tiny $-\!k_3'$}}}}
\put(5182,-822){\makebox(0,0)[lb]{\smash{\SetFigFont{9}{10.8}{\rmdefault}{\mddefault}{\updefault}{\tiny $\!k_3'\!\!-\!\!k_2'$}}}}
\put(4672,-822){\makebox(0,0)[lb]{\smash{\SetFigFont{9}{10.8}{\rmdefault}{\mddefault}{\updefault}{\tiny $\!k_2'\!\!-\!\!k_1'$}}}}
\put(4163,-822){\makebox(0,0)[lb]{\smash{\SetFigFont{9}{10.8}{\rmdefault}{\mddefault}{\updefault}{\tiny $k_1'$}}}}
\end{picture}

%% file: discussion.tx
\subsection{Discussion of the Results} \label{Sec:facres:FGDiscussion}
In the previous sub-sections we have presented our main results on the gluonic
twist-four contributions, viz.\ Eqs.\,\eqref{facres:H1gaugeInv1},
\eqref{facres:H2gaugeInv1} and \eqref{facres:H3gaugeInv1}.  While the color
structure of the expressions for $H_1$ and $H_2$ is valid beyond one
loop\footnote{The Ward identities derived in section~\ref{Sec:Ward} are valid
  only for 1PI vertex functions.  Therefore, contributions of diagrams which
  involve three- and four-gluon vertices like those in
  Fig.\,\ref{ThreeFourGVert1} have to be incorporated separately into the
  expressions for $H_1$ through $H_3$.}, we have some restrictions in the case
of $H_3$.  For a four-gluon system in the color singlet, there are five
linearly independent states in the case of even C-parity. For example, one can
expand the color structure with the help of the five states $1$, $8_\mathrm A$,
$8_\mathrm S$, $10+\overline{10}$ and $27$.  However, a closer inspection of
our ansatz Eq.\,\eqref{ward:FourGcolorDecomp1} for the color structure of the
four-gluon quark-loop shows that the state $10+\overline{10}$ does not occur
and there are, in addition to the state $8_\mathrm A$, only two specific linear
combinations of the three states $1$, $8_\mathrm S$, and $27$. In an analysis
valid to all orders one should expect that all color states contribute.

Let us, now, compare our results with those obtained by EFP \cite{biblio:EFP}.
As mentioned in the introduction, the computations in \cite{biblio:EFP} can be
divided into two steps. In the first step one considers the relevant diagrams
and performs the factorization of the coefficient functions and matrix elements
of the relevant operators.  This leads to the operators in
Eqs.\,\eqref{intro:EFPTwoQuarkOp1}-\eqref{intro:EFPFourQuarkOp1}.  They can be
compared with the two- and four-gluon operators in
Eqs.\,\eqref{facres:G2matrixExp1}, \eqref{facres:G3matrixExp1} and
\eqref{facres:G41}, which involve the same number of covariant derivatives as
the corresponding quark operators.

In the second step, EFP used equations of motion in order to express the
contributions of the two-quark operators
\eqref{intro:EFPTwoQuarkOp1}-\eqref{intro:EFPTwoQuarkOp3} in terms of linearly
independent quantities.  Within the transverse basis, the resulting twist-four
contributions to $F_\CLrT$ and $F_\CLrL$ are given in
Eqs.\,\eqref{intro:EFPSolutionFT} and \eqref{intro:EFPSolutionFL} plus terms
which involve the four-quark operator \eqref{intro:EFPFourQuarkOp1}. In our
case this second step has not yet been performed. It cannot be done without an
explicit computation of the quark-loops $R$.

On the other hand, one should try to find the relation between our results and
the operators \eqref{intro:Two_G_Op1} and \eqref{intro:Four_G_Op1} occuring in
the OPE. For this, one has, again, to compute the expressions for the
quark-loops $R$.  As has been described in \cite{biblio:EFP}, the connection to
the operators \eqref{intro:Two_G_Op1} and \eqref{intro:Four_G_Op1} can be
established for moments of contributions to the structure functions with
definite signature only, i.e.\ the expressions for $_3G^{\mu\sigma\nu}$ and
$_4G^{\mu\sigma\lambda\nu}_i$ in \eqref{facres:G3matrixExp1} and
\eqref{facres:G41} have to be separated into symmetric and anti-symmetric
parts.  Note that the Lorentz-indices of the operators in
\eqref{facres:G3matrixExp1} and \eqref{facres:G41} are not symmetrized.  The
permutations involved in Eqs.\,\eqref{facres:H2gaugeInv1} and
\eqref{facres:H3gaugeInv1} imply the simultaneous exchange of Lorentz-indices
of the quark-loop expressions and of the operators (cf.\ 
\eqref{dimana:Tensor3G1} and \eqref{dimana:Tensor4G1}).  Taking moments of
Eqs.\,\eqref{facres:H1gaugeInv1}, \eqref{facres:H2gaugeInv1} and
\eqref{facres:H3gaugeInv1} will then result in an expansion of the operators in
\eqref{facres:G2matrixExp1}, \eqref{facres:G3matrixExp1} and \eqref{facres:G41}
about the origin, so that finally the operators \eqref{intro:Two_G_Op1} and
\eqref{intro:Four_G_Op1} will be recovered.  In summary, the approach used here
is rather intuitive since one starts with certain diagrams which yield
properties of the coefficient functions which, in turn, imply restrictions on
the contributing operators. Within the OPE, on the other hand, one starts with
a certain class of operators and has to find the corresponding coefficient
functions.

As mentioned before, the computation of the quark-loop expressions $R$ will
yield terms which are proportional to logarithms in $Q^2$ and terms which are
just constants.  The terms which are proportional to logarithms have, actually,
to be interpreted as part of the evolution of the contributions in
Eqs.\,\eqref{intro:EFPcontribs1} and \eqref{intro:EFPcontribs2}.  They are due
to configurations with collinear momenta of quarks and gluons and can be
interpreted as contributions where one of the gluons splits up into a
quark-antiquark pair.  Non-logarithmic terms, on the other hand, correspond to
the NLO twist-four gluon coefficient functions. For the computation of the
quark-loop expressions $R$ it will, therefore, be necessary to re-identify the
diagrams considered by EFP\@.


%% file: conclu.tx
\section{Conclusions} \label{Sec:Conclu}
For the description of deep inelastic structure functions in the low-$Q^2$
regime, the determination of the higher-twist contributions remains one of the
outstanding theoretical challenges. At least in the small-$\CLxB$ domain, some
of these higher-twist contributions are known to increase faster with
decreasing $\CLxB$ than leading-twist terms.  In addition, their
$1/Q^2$-suppression is weakened due to evolution which is stronger than in the
leading-twist case \cite{biblio:BarBon}.  In this paper we have presented
results on gluonic twist-four operators, which are expected to give the main
twist-four contributions at small-$\CLxB$.  For the derivation of these results
we have extended the method developed by EFP \cite{biblio:EFP}. Based on the
diagrams in Fig.\,\ref{FactDiags1} and working in the axial gauge we could
perform the factorization of the upper quark-loop parts and the lower blobs.
The latter are expressed as matrix-elements of gluonic operators between proton
states. This factorization procedure requires the use of Ward identities, which
have been derived in section~\ref{Sec:Ward}. The axial gauge proves to be very
convenient for two reasons. First, gluonic $A$-operators occurring between
proton states can easily be expressed in terms of $F$-operators which is
necessary in order to obtain gauge invariant expressions
(section~\ref{Sec:Ops}).  Secondly, diagrams with more than four $t$-channel
gluons lead to contributions of, at least, twist-six.  This statement is proved
in section~\ref{Sec:DimAnalysis}.  Working in a covariant gauge, one would have
to take into account contributions of diagrams with arbitrarily many
$t$-channel gluons.

After the factorization had been performed we ended up with three
contributions.  The two- and three-gluon quark-loops are convoluted with matrix
elements of the two-gluon operator, Eqs.\,\eqref{facres:H1gaugeInv1} and
\eqref{facres:H2gaugeInv1}, while the four-gluon quark-loop is convoluted with
both, the two- and the four-gluon operator, Eq.\,\eqref{facres:H3gaugeInv1}.
In order to derive the latter result, we had to make intensive use of symmetry
relations for the expression of the four-gluon quark-loop. These symmetry
relations also led to restrictions on the color structure of the four-gluon
operator contribution.  As the next step we have to define a minimal set of
operators in terms of which we can express the operators in
\eqref{facres:H1gaugeInv1}, \eqref{facres:H2gaugeInv1} and
\eqref{facres:H3gaugeInv1}. This requires the introduction of signature of the
two- and four-gluon $t$-channel states, i.e.\ a further step is symmetrization
of the color and Lorentz indices. It is only after this decomposition into
contributions with definite signature that we can return to
\eqref{intro:Two_G_Op1} and \eqref{intro:Four_G_Op1} and determine
contributions to the structure functions.
 
It is worthwhile to compare the results presented here with some earlier work.
The traditional approach to small-$\CLxB$ physics is the high-energy
(small-$\CLxB$) approximation, which, in leading order, leads to the BFKL
equation \cite{biblio:BFKL}. The BFKL equation is usually described by a ladder
diagram of two reggeized $t$-channel gluons. It sums all leading logarithms in
$\CLxB$ and contains contributions of \textit{all} twists.  Contributions of
diagrams with three and four reggeized $t$-channel gluons have been studied in
the generalized leading logarithmic approximation in
\cite{biblio:Bartels,biblio:BarWue}\,\footnote{Recently, C.\,Ewerz
  \cite{biblio:EwerzPhD} has performed an analysis of the five- and six-gluon
  amplitudes.}.  The results of these calculations have been used
\cite{biblio:BarBon} to compute the twist-four part in the double-logarithmic
approximation (DLA), and it is interesting to compare them with the results of
the present paper.

First, in \cite{biblio:BarWue} it was shown that the three-gluon amplitude can
be written as a sum of expressions for the two-gluon (BFKL) amplitude. Since
this phenomenon generalizes the reggeization of the gluon, it has been termed
reggeization, as well.  Color structure and reggeization are independent of the
twist. The fact that the three-gluon amplitude can be written as a sum of
two-gluon amplitudes is therefore also true for the twist-four part. This
result can be confronted with our new result that the three-gluon quark-loop
couples only to the two-gluon operator \eqref{facres:H2gaugeInv1}. Next, the
four-gluon amplitude studied in \cite{biblio:BarWue} splits up into two terms:
there is, again, a reggeizing part, which can be expressed with the help of
expressions for the two-gluon amplitude, and a second part which is irreducible
with respect to reggeization. We are, now, tempted to draw a relation with
Eq.\,\eqref{facres:H3gaugeInv1} which also contains contributions from
two-gluon and four-gluon operators.  In particular, one might associate the
second matrix element in \eqref{facres:H3gaugeInv1} with the reggeizing part of
\cite{biblio:BarWue}, whereas the first term in \eqref{facres:H3gaugeInv1}
looks analogous to the irreducible four-gluon amplitude.  Here, however, a word
of caution is in place. First, in \cite{biblio:BarWue} it is the reggeizing
part, which contains color factors like $d^{abcd}$, while the irreducible part
is symmetric under simultaneous exchange of color and momentum. This is in
contrast to the color structure of two- and four-gluon operators in
\eqref{facres:H3gaugeInv1}. Secondly, the calculations use different
approximations. Ref.\ \cite{biblio:BarWue} studies the leading-$\log(1/\CLxB)$
approximation. This has lead to the result that the lowest-order diagram with
four $t$-channel gluons belongs to the reggeizing part, which suggest that
there is no direct coupling of the four-gluon operator to the quark loop. In
the present paper we have taken a different route: our analyis of $1/Q^2$
corrections has not been restricted to the small-$\CLxB$ approximation, and we
find contributions of the two- and four-gluon operator already at
leading-order.  This can be considered as a hint that one will encounter
simplifications if one uses explicit expressions for the four-gluon quark loop
in those terms which are coupled to the four-gluon operator in
Eq.\,\eqref{facres:H3gaugeInv1} and then takes the limit $\CLxB \rightarrow 0$.
Agreement between the results of \cite{biblio:BarWue} and the present approach
holds only in DLA; to go beyond this approximation is one of the main
motivations of the present paper.

In summary, we have performed the first step towards a consistent treatment of
gluonic twist-four contributions at small-$\CLxB$. What remains to be done is,
first of all, the computation of the expressions for the quark-loops.  While
this task might be straightforward, one has to keep in mind that the
expressions will contain contributions proportional to $\log Q^2$ and,
therefore, contribute to leading-order quark-coefficient functions convoluted
by Altarelli-Parisi splitting functions as well as to next-to-leading order
gluon coefficient functions.  A major task will be the computation of anomalous
dimensions of the gluon operators. Since several gluonic operators contribute
to twist-four, one will have to handle also the problem of operator mixing.
First results are contained in \cite{biblio:BarBon} and
\cite{biblio:Bukhvostov}, but a complete LO analysis does not exist.  Moreover,
although we have argued that at small-$\CLxB$ gluon-operators are expected to
dominate, the r\^ole of quark-operators has not been clarified. For a fully
consistent treatment of twist-four contributions one should also take
contributions of quark-operators into account. In such an analysis one will
encounter the problem of operator-mixing, as well.  We feel that calculations
along these lines are urgently needed: a more complete understanding of deep
inelastic $ep$ scattering at low $Q^2$ and small $\CLxB$ cannot be reached
unless we know the influence of higher twist.

{\bf Acknowledgements:} We wish to thank L.N.\,Lipatov and R.K.\,Ellis for
stimulating and very helpful discussions.